\documentclass[a4paper,12pt]{article}
\usepackage{amsmath,amsthm,amsfonts,setspace,enumerate,graphicx,tikz,placeins,hyperref,bbm,markdown}
\usetikzlibrary{patterns,decorations.pathreplacing,decorations.markings}
\usepackage[margin=1in]{geometry}
\usepackage[title]{appendix}

\usepackage[font=small,labelfont=bf]{caption}
\newenvironment{dense}{
 \medmuskip=0mu
 \thinmuskip=0mu
 \thickmuskip=0mu}

\tikzset{every picture/.style={line width=0.75pt}} 

\tikzset{
  pattern size/.store in=\mcSize,     pattern size = 5pt,
  pattern thickness/.store in=\mcThickness,  pattern thickness = 0.3pt,
  pattern radius/.store in=\mcRadius, pattern radius = 1pt
}

\makeatletter
\pgfdeclarepatternformonly[\mcThickness,\mcSize]{diagBlack}%
{\pgfqpoint{0pt}{-\mcThickness}}{\pgfqpoint{\mcSize}{\mcSize}}{\pgfqpoint{\mcSize}{\mcSize}}{%
  \pgfsetcolor{\tikz@pattern@color}%
  \pgfsetlinewidth{\mcThickness}%
  \pgfpathmoveto{\pgfqpoint{0pt}{\mcSize}}%
  \pgfpathlineto{\pgfqpoint{\mcSize+\mcThickness}{-\mcThickness}}%
  \pgfusepath{stroke}%
}
\pgfdeclarepatternformonly[\mcThickness,\mcSize]{diagGray}%
{\pgfqpoint{0pt}{-\mcThickness}}{\pgfqpoint{\mcSize}{\mcSize}}{\pgfqpoint{\mcSize}{\mcSize}}{%
  \pgfsetcolor{\tikz@pattern@color}%
  \pgfsetlinewidth{\mcThickness}%
  \pgfpathmoveto{\pgfqpoint{0pt}{\mcSize}}%
  \pgfpathlineto{\pgfqpoint{\mcSize+\mcThickness}{-\mcThickness}}%
  \pgfusepath{stroke}%
}

\usepackage{natbib}
\usepackage[T1]{fontenc}

\hypersetup{
 colorlinks = true,
 linkcolor=blue,
 citecolor=blue,
 urlcolor=blue,
 filecolor=blue,
 bookmarksnumbered, 
 bookmarksopen=true, 
 bookmarksopenlevel=1,
}

\renewcommand{\t}{\gamma}

\newcounter{thm}
\newcounter{lm}
\newcounter{cj}
\newcounter{pr}

\newtheorem{theorem}[thm]{Theorem}
\newtheorem{lemma}[lm]{Lemma}

\newtheorem{corollary}[cj]{Corollary}
\newtheorem{proposition}[pr]{Proposition}
\newtheorem{remark}{Remark}

\def\a{\alpha}

\def\g{\gamma}

\def\e{\varepsilon}

\def\h{\eta}
\def\th{\theta}

\def\l{\lambda}

\def\s{\sigma}
\def\t{\tau}

\label{key}

\def\D{\Delta}

\def\GG{\mathcal{G}}

\def\UU{\mathcal{U}}

\def\del{\partial}

\DeclareMathOperator{\E}{\mathbb{E}}
\DeclareMathOperator{\real}{\mathbb{R}}

\newcommand{\de}{\mathop{}\!\mathrm{d}}

\newcommand{\Paren}[1]{\left( #1 \right)}

\newcommand{\brac}[1]{[ #1 ]} 

\newcommand{\Brac}[1]{\left[ #1 \right]}

\title{The Tension between Trust and Oversight in Long-term Relationships\thanks{We thank Martin Cripps, Jack Fanning, Daniel Hauser, and audiences at EWET 23 and EWMES 23 for helpful comments and suggestions. Knoepfle acknowledges financial support by the Academy of Finland (project number 325218).}}
\author{Peter Achim\footnote{Dept. of Economics, University of York.  peter.wagner@york.ac.uk.}
\and 
Jan Knoepfle\footnote{School of Economics and Finance, Queen Mary University. j.knoepfle@qmul.ac.uk.}}

\begin{document}
\begin{titlepage} 

\maketitle \thispagestyle{empty} 

\begin{abstract}
A principal continually decides whether to approve resource allocations to an agent, who exerts private effort to remain eligible. The principal must perform costly inspections to determine the agent's eligibility. We fully characterize Markov Perfect Equilibria and analyze the resulting paths of trust and oversight that emerge from the dynamic interplay of effort and oversight incentives. At high trust levels, effort is an intertemporal substitute to oversight, which leads to unique interior effort choices and random inspections. At low trust levels,  effort is an  intertemporal complement to oversight, which may create a coordination problem, leading to equilibrium multiplicity. Voluntary disclosure can mitigate this coordination issue.
\end{abstract}

\noindent Keywords:  Oversight dynamics, Reputation,  Inspections
\\
\noindent JEL: C73, D83, F35

\end{titlepage}

\section{Introduction}

A balance of trust and oversight is essential for the success of relationships that operate at a distance. Trust is a conduit for cooperation, reduces the need for oversight, and facilitates transactions that would otherwise not be mutually beneficial \citep{kreps_1990}. Foreign aid relations, for example, are possible only if donor countries place trust in the recipient country’s ability to manage corruption and invest resources effectively. International supply chains depend upon the offshoring client’s trust that the overseas supplier adheres to manufacturing standards and worker safety protocols. Trust is similarly foundational for international relations, or the contracting of public service providers. 
Trust cannot exist without some level of oversight, as trust is based on reputation and built over time through repeated positive encounters \citep{Dasgupta1988}. 
Trust and oversight thus exhibit a natural tension: if trust reduces the need for oversight but cannot replace it, how do trust and oversight evolve and interact in long-term relationships?

The literature following \cite{kreps1982reputation} has explored the notion of trust as the reputation of a long-lived player interacting with a sequence of short-lived players \citep[for a review, see][]{Mailath&Samuelson2001}. We introduce a novel perspective on the dynamics of oversight and its interplay with trust. Endogenizing the monitoring choices of \textit{short-lived} players, \cite{liu2011information} highlights the cyclical nature of trust, where reputation is repeatedly built and ``milked.'' Our work examines the dynamics of oversight exerted by a \textit{long-lived}, forward-looking principal. We show that oversight, much like trust, undergoes cycles: after each performed inspection, there is an oversight break. The structure of oversight behavior after the break crucially depends on the level of trust. 

Our model incorporates the capital-theoretic approach to reputations of \cite{board2013reputation} into a principal-agent game. The agent builds reputation over time by investing effort in partially persistent quality. Quality may be understood literally as the quality of a provided service, or, as in the case of foreign aid, as the recipient's compliance with agreed conditions, like measures to prevent fund embezzlement. The principal decides whether to approve resource allocations based on her trust in the agent, which is informed by past inspections of quality. In the applications above, approval means that foreign aid is paid out, that a manufacturer continues using her offshore supplier, or that the government continues the service contract with a  utilities provider. 

We assume that quality is non-contractible and the principal has no commitment power. The absence of commitment reflects the fact that in many applications, incentives are informal rather than supported by explicit formal contracts. \cite{bourguignon2020foreign} cites the lack of commitment as a key challenge in foreign aid relations, where donors tend to refrain from withholding funds even though conditions for aid imposed on recipients are not met.\footnote{This phenomenon is also referred to as the ``Samaritan's Dilemma''  \citep{svensson2000}.} To isolate the role of trust in the relationship, we focus on equilibria in Markov strategies, which depend on the principal's level of trust. This results in a sharp characterization of the dynamic interaction between trust and oversight.

\paragraph{Results.} The equilibrium characterization in Theorem \ref{thm:EQ characterization} shows that trust and oversight jointly undergo cycles. Each  inspection is followed by a break with no oversight. High trust is milked by shirking, and the principal inspects at random after the break. Low trust can prompt recovery, where the agent exerts full effort and the principal inspects deterministically after the break. 

These features match observed inspection patterns in multiple real-life settings: The US Occupational Health and Safety Administration (OSHA) specifies a minimum break of two years between consecutive inspections \citep{ko2010role}. 
Cyclical patterns of milking and recovery are frequently observed in foreign aid. For the recipient countries Pakistan \citep{mccartney2012competitiveness} and Kenya \citep{deaton2013great}, for instance, these cycles were documented to involve rising corruption, leading to a freeze in donor funds, prompting the recipients to enact reforms which eventually pave the way for a resumption of aid.

Equilibria characterized in Theorem \ref{thm:EQ characterization} entail three different regions, determined by the level of trust. At high levels of trust, the principal grants approval without oversight. In this  \textit{blind trust} region, the agent exerts no effort and trust gradually declines. 
At low levels of trust, the principal grants no approval and performs no oversight. In this \textit{blind distrust} region, whenever effort can be induced, the equilibrium behavior is not unique. 
Low trust can always result in a breakdown of the relationship: the agent exerts zero effort and the principal never inspects again. Low trust can also prompt recovery: the agent exerts full effort and trust gradually increases. 
Intermediate levels of trust form the oversight region. 
The inspections timing is determined by the level of trust within this oversight region. If trust is high initially and declines toward the oversight region, the principal inspects randomly at a constant rate. Conversely, if trust is low initially and increases toward the oversight region, the principal inspects deterministically as soon as the oversight region is reached.  

 The different inspection-timing structures  stem from the intertemporal strategic effects between effort and oversight across different levels of trust.\footnote{To the best of our knowledge \cite{jun2004strategic} are the first to analyze the distinction between contemporaneous and intertemporal strategic substitutes/complements. They show for a dynamic duopoly model that contemporaneous strategic substitutes can turn to intertemporal strategic complements. We find similar effects at low levels of trust.} Oversight is a (contemporaneous and intertemporal) \textit{complement} for effort at all levels of trust: an increase in current or future inspection probabilities generates more effort incentives for the agent. For the principal, on the other hand, effort is an intertemporal \textit{substitute} for oversight at high levels of trust: a reduction in the agent's cumulative effort increases inspection incentives. The incentives of the agent and the principal thus work in opposite directions. Therefore, at high trust, the equilibrium involves an ``interior solution'' with random inspections and interior effort. In contrast, effort is an intertemporal \textit{complement} to oversight at low levels of trust. At low trust, principal and agent face a form of coordination problem. Equilibrium involves a ``corner solution'' and multiplicity arises:
 the parties either end the relationship with zero effort and no oversight, or they recover trust through a period of full effort and a deterministic inspection.  
 
 Theorem \ref{thm:multiplicity of EQ} presents existence conditions for the different types of equilibria.  Remarkably, when oversight becomes cheaper, the principal's increased incentive to inspect can eliminate equilibria with positive effort. That is, cheap oversight can hurt the relationship as excessive oversight lowers the value the agent derives from the relationship and thereby reduces his effort incentives. 

Our analysis highlights that trust relationships with costly oversight can be mutually beneficial despite fluctuations and recurring decline. Indeed, an important determinant for the value of the relationship is the ability to recover trust efficiently. As a consequence of the differing strategic effects, communication from the agent to the principal can improve efficiency during low trust (Theorem \ref{thm:self-disclosure}) by resolving the coordination problem,\footnote{A frequent problem in aid programs is achieving time consistency. Recipient countries have an incentive to cease reform efforts once aid flows have been interrupted \citep[see][]{collier1997redesigning,bourguignon2020foreign}. This can lead to a breakdown of the aid relationship. Our results suggest that communication channels are particularly valuable in reducing the prevalence of inefficient breakdowns.} but there is no gain from communication at high trust because interests are too opposed.
 
The rest of the paper is organized as follows. After the discussion of the related literature, Section \ref{sec:Model} sets up the model. Section \ref{sec:Equilibrium} analyzes the agent's and the principal's optimization problems and presents the equilibrium characterization. Sections \ref{sec:Multiplicity} and \ref{sec:Self-disclosure}  explore the coordination problem in depth, showing that cheap oversight can be detrimental, and how communication can foster efficiency. Section \ref{sec:conlusion}  concludes.

\paragraph{Literature.} 
To study the tension between trust and oversight, we introduce costly inspections into a reputation setting. Conceptually, the closest paper is \cite{liu2011information}, which introduces endogenous information acquisition by short-lived agents into a reputation model of the ``adverse selection approach'', where short-lived players are uncertain about the long-lived player's type.\footnote{See \cite{mailath2006repeated} p. 459 for a discussion. For adverse-selection reputation models with two (equally patient) long-lived players, see e.g. \cite{atakan2012reputation} and \cite{atakan2015reputation}.} 
\cite{liu2011information} enhances the applicability of reputation models and offers an explanation, for example, for the endogenous emergence of reputation cycles. Since the monitoring agents are short-lived and identical, this model cannot speak to the \emph{dynamics} of oversight -- players randomize identically in each period. In our model a long-lived principal chooses when to inspect, resulting in rich oversight dynamics, evolving in cycles with each inspection followed by a grace period with no oversight. 

Our model builds on the capital-theoretic model of reputation by \cite{board2013reputation}. In their model an agent operates in a market which learns about his partially persistent quality through an exogenous arrival process. Several papers consider variations of this model in which the arrival of information is endogenous. \cite{hauser2023censorship, hauser2023promoting} considers models in which the agent manipulates the rate at which information is disclosed through censorship of bad or promotion of good news, respectively. Our approach is closer to the self-certification model of \cite{marinovic2018dynamic} in which the agent controls the exact timing of information disclosure to the market.\footnote{See also \cite{pei2016reputation} for a model of strategic disclosure under reputation concerns.} The crucial difference is that in our principal-agent setting, it is the principal as receiver who chooses to acquire information through costly inspections. This separation of effort choices and information acquisition has important implications for the equilibrium dynamics, and forms the foundation for the tension between trust and oversight that is the focus of our paper. In contrast to
\cite{marinovic2018dynamic}, equilibria with positive effort in our framework require the use of mixed inspection strategies. 

A small number of papers study inspections by embedding the \cite{board2013reputation} reputation framework to a principal-agent setting. \cite{Varas2017} consider a pure moral-hazard model in which the agent is motivated by the desire to maintain a good reputation and inspections make the agent's type public. The authors show that with linear payoffs, the principal would optimally commit to inspect randomly at a constant rate. In \cite{achim2024relational}, we consider an enforcement setting with transfers, where the principal uses fines and inspections to induce maximum compliance of the agent. In this equilibrium, the agent honestly discloses his private information at all times, and thus reputational concerns play no role. In contrast, in the present paper there is no money, which makes it impossible for the principal to induce full effort or honest reporting by the agent. This makes reputational concerns central and enables us to study dynamic paths of trust over time that result from the intertemporal interplay between effort, reputation and oversight. 
Another important difference  in the present paper is the focus on Markov strategies. The principal cannot commit or be incentivized to inspect through coordinated continuation play, so that her incentive to inspect stems purely from the desire to acquire information.

The absence of money relates our paper to the growing literature on dynamic incentive problems in the absence of money. A consistent finding in this literature is that the relationship tends to deteriorate over time. \cite{horner2015dynamic} study optimal dynamic mechanisms when the principal can commit to use future allocations to elicit private information. \cite{li2017power} considers a repeated game of project selection in which a principal relies on the advice of a biased agent and offers power over future allocation decision to incentivize the agent. 
Both papers find that the optimal relationship features increasingly inefficient allocations. 
\cite{lipnowski2020repeated} consider a repeated allocation problem without commitment and also find that the agent increasingly acts against the principal's interest, and the principal in turn rarely delegates. In our paper, in contrast, the relationship is not doomed to failure, but may  evolve in cycles. Recovery and breakdown can coexist in our model because of the intertemporal strategic complementarity between effort and oversight at low levels of trust.

Our paper is related to the wider literature on optimal monitoring.\footnote{For dynamic contracts with commitment power, see \cite{Antinolfi2015, piskorski2016optimal, chen2020optimal, li2020optimal,solan2021dynamic,dai2022dynamic, wong2022dynamic,solan2023not} for monitoring of the current action.
For dynamic inspections of a partially persistent state see \cite{ravikumar2012optimal,Kim2015,Varas2017,ball2023should,achim2024relational}.} Most closely related within this literature are papers that focus on the incentives of the monitor. \cite{strausz1997delegation} shows in a static model, that restricting a principal's commitment limits her use of monetary incentives and increases the benefits of delegating the monitoring task.
In \cite{halac2016managerial} and \cite{dilme2019residual}, the principal invests to adjust her persistent monitoring capabilities. In both papers, the principal wants to catch agents who misbehave as the only way to signal vigilance. In that regard, the principal's incentive to inspect in these papers is largest when she expects the agent to misbehave in this instant. In our paper, inspection incentives crucially depend on the expected cumulative effort of the agent, and the effect on oversight incentives goes in either direction, depending on the current level of trust. 

Finally, our model bears resemblance to classic delegation frameworks as we could interpret the principal's approval as delegating decision authority to a the (biased) agent. \cite{alonso2008optimal} propose a model where the principal commits ex ante to a delegation rule that restricts the agent’s discretion so that his actions conform to the principal’s interests. \cite{amador2013theory} consider a similar problem by setting fixed bounds on the agent’s decisions to mitigate information asymmetries. In contrast to these papers, our model is dynamic and the principal has no commitment power. As a result, the focus of our analysis is not the design of optimal delegation mechanisms ---the principal will grant approval whenever myopically optimal--- but rather to uncover the intertemporal trade-offs and the sustainability of long-run relationships which rely on costly oversight, an aspect that static delegation models that treat the state as exogenous and allow for commitment in the delegation rule do not address.

\section{Model}\label{sec:Model}
\paragraph{Players, actions, and state dynamics.} 
There are an agent and a principal. Time $t \in [0, \infty)$ is continuous. The agent, at each time~$t$, chooses effort $\eta_t \in [0,1]$ to invest in quality. 
Quality follows a two-state Markov process with $\theta_t \in \{0,1\}$ starting at $\th_0 = 1$. Following \cite{board2013reputation}, quality transitions from 1 to 0 at Poisson rate $ \l(1- \eta_t)$, and from 0 to 1 at rate $\lambda \eta_t $. We may interpret $\lambda$ as the arrival rate of shocks such that, if there is a shock at time~$t$, the resulting state is $\theta_t=1$ with probability $\eta_t$ and it is $\theta_t=0$ with probability $1-\eta_t$. The state is constant between shocks.

The principal chooses approval and inspections. At each instant $t$, she makes an approval decision $\alpha_t \in \{0,1\}$. We say that the principal \textit{trusts} the agent at time~$t$ if $\alpha_t =1$, and she \textit{distrusts} the agent if $\alpha_t =0$. 
The only way for the principal to learn $\th_t$ at times $t>0$ is by performing an inspection at lump-sum cost $k$. 
We assume that the principal cannot commit to future approval or inspection decisions.\footnote{If the principal could commit, the optimal policy would be stationary in the long run. The principal would either inspect at a fixed rate and always approve while the agent always works, or the principal would never inspect nor approve and the agent never work. Unsurprisingly, the principal would be better off with commitment. Note that a principal with commitment power would inspect despite knowing the state. In contrast, we want to focus on inspection activity driven by the desire to learn the state.}

\paragraph{Payoffs.} Both players are risk-neutral and discount future payoffs at common rate $r > 0$.
The agent's total expected discounted payoff given a pair of measurable approval and effort strategies  $\{\alpha_t\}_{t\ge 0}$ and $\{\eta_t\}_{t\ge 0}$ is 
\begin{align*}
\E \Brac{\int_0^\infty e^{- r t}\Paren{\a_t u - \eta_t c}\de t}.
\end{align*} 
That is, he gains flow utility $u$ from approval and faces marginal effort cost $c>0$. Assume $u/(r+\l) >c/r$; otherwise the agent would never exert effort even if $\theta_t$ was always observable by the principal.

The principal's total expected discounted payoff under approval strategy $\{\alpha_t\}_{t\ge 0}$ and inspection times $\tau_1 < \tau_2 <\cdots$ is
\begin{align*}
\E\Brac{\int_0^\infty e^{- r t}\a_t v(\theta_t)\de t-   \sum_{i=1}^\infty e^{-r \t_i} k},
\end{align*}
with $v(\th) = \th H - (1-\th) L$. That is, her flow payoff while approving ($\a_t=1$) is $H>0$ in state 1 and $-L<0$ in state $0$. The flow payoff during disapproval ($\alpha_t=0$) is normalized to zero. The principal pays lump-sum cost $k>0$ each time she performs an inspection.

\paragraph{Reputation Dynamics and Markov Strategies.}
A public history $h_t$ consists of all inspection times $\t_i$ and inspection outcomes $\theta_{\t_i}$ up to and including time $t$.\footnote{As in \cite{board2013reputation}, for any principal-strategy, the agent's effort incentives will be independent of the true state. Therefore, for the core analysis it is irrelevant if the agent observes $\theta_t$ at all times or not. Section~\ref{sec:Self-disclosure} contains an extension where the agent observes the state can report it via cheap-talk messages. Further, given our focus on Markov strategies and since  approval has no impact on state transitions, recording $\{\alpha_s\}_{s\le t}$ as part of the history would be inconsequential.} 
Denote the agent's \textit{reputation} by $p_t = \E[\theta_t \lvert h_t ]$, i.e. the expected quality based on the public information available at time~$t$ and the anticipation of the agent's (unobserved) effort strategy. 
By the Law of Iterated Expectations\footnote{Since $v$ is linear, we have $\E\Brac{v(\theta_t )} = \E\Brac{\E[v(\theta_t) \lvert h_t]} = \E\Brac{v(\E[\theta_t \lvert h_t])} = \E\Brac{v(p_t)}$ for all $t$.} we can rewrite the principal's discounted expected payoff as 
\begin{align}\label{eq:payoff principal with p}
\E\Brac{\int_0^\infty e^{- r t }\a_t v(p_t)\de t - \sum_{i=1}^\infty e^{-r \t_i }k  },
\end{align}
with the natural extension of value $v$ to $[0,1]$ as $v(p)  = p H - (1-p) L$.

This equivalence indicates the suitability of reputation $p$ to serve as a state variable. 
To focus on the tension between trust and oversight, we study Markov perfect equilibria with reputation as the state variable.\footnote{See \cite{achim2024relational} for the equilibrium analysis in a related model, but without the Markov restriction. In that case, inspection incentives can be created through coordinated continuation play rather than the principal's desire to learn the state.}

 \begin{remark}
To ensure that the principal cannot learn the state without inspection, we assume she does not observe the payoff per state $v(\th_t)\in \{-L,H\}$. 
This assumption is common in  repeated-games  with imperfect monitoring, and often justified by assuming the game ends at rate $r$ and accumulated payoffs are collected at the end. 
We can allow the principal to observe the time-$t$ expectation $v(p_t) \in [-L,H]$.  
Imperfect observability is realistic in environments with informational distance due to temporal, spatial or institutional separation. For example, many corporations benefit from offshore suppliers’ compliance with labor standards without continually observing exact working conditions.
Foreign aid providers value recipient countries’ anti-corruption efforts without being affected by them in real time. 
Alternatively, and closer to \cite{board2013reputation},  the true payoff may be determined by $p_t$ directly. For example, the political benefits of approving foreign aid provision may depend on the own electorate's expectation of $\theta_t$.
 \end{remark}

A Markov strategy for the agent $\eta\colon [0,1]\to [0,1]$ specifies the effort level $\eta_t = \eta(p_{t-})$, where $p_{t-}= \lim_{s\nearrow t}p_s$ denotes the left limit of the reputation at time $t$.\footnote{Since reputation paths $\{p_t\}_{t\ge 0}$ are continuous almost everywhere, $\eta(p_{t-}) = \eta(p_{t})$ for almost all $t$. Further, since the probability of a transition shock is zero for any isolated $t$, this distinction is immaterial.}
It will be convenient to explicitly distinguish the (Markov) strategy $\tilde{\eta}$ that the principal believes the agent uses. Naturally, $\tilde \eta =\eta$ in equilibrium. 
If the principal does perform an inspection at time $t$, then reputation is updated gradually via Bayes' rule according to $\tilde \eta$.  
Thus, the evolution between inspections follows the ordinary differential equation (ODE)
\begin{align}\label{eq:ReputationODE}
\de p_t=  \lambda (\tilde\eta(p_t)-p_t )\de t.
\end{align}
If the principal performs an inspection at any time $\t$, the true state is revealed and the reputation jumps to $p_\t = \th_\t$. 
Note that certain believed strategies $\tilde{\eta}(\cdot)$ may admit no solution to the differential equation. To resolve this, we follow \cite{board2013reputation} and impose restrictions on $\tilde \eta$, which are given in  Appendix~\ref{subsec:BeliefRestriction}.

For the principal, a Markov approval strategy $\alpha\colon [0,1]\to \{0,1\}$ assigns approval $\alpha_t = \alpha(p_t)$ according to the current reputation.
The principal's Markov inspection strategy $\s\colon [0,1] \to \real_+ \cup \ \{\infty\}$, gives the hazard rate $\s(p_{t-})$ of an inspection at time $t$ according to the reputation just before $t$. We denote by $\eta(p)=\infty$ the event that the principal inspects immediately with probability 1 when the reputation reaches $p$. Note that this definition rules out inspections being performed with positive probability (mass points) strictly in $(0,1)$. We allow for this possibility when analyzing the principal's best responses and find that there is always an outcome equivalent equilibrium in which all inspections arrive with finite hazard rate or immediately\footnote{This stems from the evolution of the principal's inspection incentives and stands in contrast to some optimal (commitment) policies found in \cite{Varas2017} and \cite{ball2023should}.} (see proof of Theorem \ref{thm:EQ characterization} in Appendix \ref{App:main proofs}).

\paragraph{Sequential Rationality.} To specify a notion of sequential rationality, we adopt a recursive formulation of payoffs over infinitesimal intervals, decomposing each player's payoff into the immediate payoff component and the discounted continuation value.

For the agent's problem, take a principal-strategy and a believed effort strategy  as given. Let $U(p,\theta)$ be the agent's continuation value function at reputation $p$ and quality $\th$. 
For any reputation $p$ that leads to an immediate inspection, the true state is revealed and the reputation jumps to $p=\th$. 
Thus, the value satisfies
\begin{align*}
    U(p,\theta) = U(\theta, \theta), \qquad \text{ for all } p \text{ with }  \s(p) = \infty.
\end{align*}
Conversely, for states  $(p,\th)$ with $\s(p)<\infty$, consider an interval $[t,t+\de t)$. 
The agent's value function $U$ must satisfy 
\begin{equation}\label{eq:U recursive}
\begin{aligned}
U(p,\th)
 = \max_{\eta \in [0,1]}\Big\{ \big[\alpha(p)\,u
 - \eta \,c
 + \lambda
 \Brac{\eta \,U (p,1) +  (1-\eta  )\,U (p,0) }  + \s(p) U (\theta,\theta) \big] \de t & \\
+  \Paren{1-r \de t -  \lambda \de t  -  \s(p) \de t  }\,
U (p+\de p,\theta)
 + o(\de t) &\Big\}.
 \end{aligned}
\end{equation}
The agent earns a flow payoff of $\alpha(p) u$ and incurs flow effort cost $\eta_t c$ given his choice $\eta_t \in [0,1]$. 
At rate $\lambda$ a potential state transition arrives, and the state jumps to $1$ with probability $\eta $ or to $0$ with probability $1-\eta$, yielding expected continuation value $\eta U(p,1) + (1-\eta) U(p,0)$.
At rate  $\s(p)$ an inspection occurs, which reveals the true state $\theta$ so that the continuation value becomes $U(\theta,\theta)$. If no transition and no inspection arrives, the reputation changes to $p+\de p$ according to \eqref{eq:ReputationODE}, and  the agent enjoys continuation value is $U(p+\de p,\theta)$ discounted with factor $ e^{-r \de t} \approx 1- r\de t $, for $\de t$ small.

For the principal's problem, fix the agent's believed effort strategy  $\tilde \eta$ and consider an interval $[t,t+\de t)$ with current reputation $p_t=p$. 
The principal's value function $V$ must thus solve the recursive equation 
\begin{equation}\label{eq:V recursive}
\begin{aligned}
 V(p)= \max\Big\{ \max_{\alpha \in [0,1] } \big\{ \alpha \,v(p) \de t
+ (1-r \de t )\,V (p+\de p )+ o(\de t) \big\} \ ; &
\\
\  p\,V (1 )+(1-p)\,V (0)- k \ 
&\Big\}.
\end{aligned}
\end{equation}
The outer maximization determines whether the principal inspects when the reputation is $p$ or not. 
In the first row, corresponding to not inspecting,
the principal earns a flow payoff of 
$\alpha v(p)$ and the discounted continuation payoff $V(p+\de p)$. 
In the second row, corresponding to an inspection, $\theta$ is revealed so that, the reputation jumps to 1 with probability $p$ and to 0 with probability $1-p$, and the principal incurs lump-sum cost $k$. If the principal mixes between inspecting and waiting, i.e. $\s(p) \in (0,\infty)$, then the  two terms must be equal.\footnote{Equivalently, we could include in the inner maximization in \eqref{eq:V recursive} the choice of the inspection rate $\s \in \real_+ \cup \ \{\infty\}$, which would add the term $\s[pV (1 )+(1-p)V (0)- k \ - V(p)]\de t$ to the first term. Naturally, this would lead to the same optimality conditions.}

\paragraph{Equilibrium Concept.} 
A Markov Perfect Equilibrium consists of strategies $\big(\a,\s,\eta,\tilde \eta \big)$
such that at any history with current reputation-quality pair $(p,\th)$, the agent's expected discounted utility is
$U(p,\th)$,  and  
the  principal's expected discounted utility is
$V(p)$; and
\begin{enumerate}[(i)]
\item The strategies are sequentially rational: conditions \eqref{eq:U recursive} and \eqref{eq:V recursive} hold. 
\item\label{item:PostInspection} At each inspection time $\t$, the  reputation updates to the true state: $p_\t = \theta_\t$.
\item\label{item:NoInspection} In the absence of an inspection, the reputation evolves according to \eqref{eq:ReputationODE}.
\item The belief about the agent's effort strategy is correct: $\tilde{\eta}=\eta$.
\end{enumerate}
 
Conditions \eqref{item:PostInspection} and \eqref{item:NoInspection} ensure that beliefs are formed according to Bayes' rule on and off the equilibrium path. In particular, \eqref{item:PostInspection} says that the reputation jumps to the true state also after the principal performs an unexpected inspection. Condition \eqref{item:NoInspection} ensures that reputation is continuously updated according to the believed effort strategy $\tilde \eta$ also after an unexpected inspection or after skipping an inspection that should have taken place with certainty.\footnote{The reputation models in \cite{marinovic2018dynamic} and \cite{hauser2023promoting} impose analogous conditions on off-path continuation play. These conditions imply the ``no-signaling-what-you-don't-know'' condition \citep{fudenberg1991perfect} that off-equilibrium actions by the principal not affect her own belief about the unknown state.}

\section{Equilibrium analysis} \label{sec:Equilibrium}
Toward the full characterization of equilibrium strategy profiles in Section~\ref{subsec:equilibrium charact}, we consider the agent's and the principal's problem separately the optimality conditions for the agent's effort, the principal's approval, and the principal's inspection strategy.
A believed effort strategy $\tilde \eta$ governs the reputation evolution between inspections. Some reputations on $[0,1]$ may never be reached, regardless of the agent’s and principal’s actions.\footnote{In particular, let $ \underline \rho = \inf \{ p \in [0,1] \ \lvert \ \tilde \eta (p) -p \le 0 \}$ and $ \bar \rho = \sup \{ p \in [0,1] \ \lvert \ \tilde \eta (p) -p \ge 0 \}$. Only reputations on $ [0, \underline \rho] \cup [\bar \rho, 1]$ can be reached.} The derivation of optimal strategies only applies to reputations not excluded by $\tilde \eta$. In Theorem~\ref{thm:EQ characterization} below, we specify Markov strategies for $p\in [0,1]$ that satisfy optimality criteria everywhere.

\subsection{The agent's problem: optimal effort choice}
\label{subsec:agent}

Following \cite{board2013reputation}, we define the \emph{value of quality} at reputation $p$ as the difference between the agent's continuation utility in the high and in the low state at a given reputation $p$:
$$D(p)= U(p,1)-U(p,0).$$ 
It is easy to see from the agent's problem in  \eqref{eq:U recursive} that $\l D(p) - c$ is the derivative with respect to current effort $\eta$. Since the reputation evolution in the future is based on believed effort $\tilde \eta$,  the value $D(p)$ fully determines the agent's optimal effort choice. 
The following result makes this precise and gives the evolution of the value of quality.

\begin{proposition}\label{prop: effort incentives}
  An effort strategy $\{\eta_t\}_{t\ge 0}$ is optimal for the agent if and only if it satisfies 
  \begin{align}\label{eq:best-response}
\eta_t
\begin{cases}
= 1 & \text{if } \lambda D(p_t)>c,\\
 =0 & \text{if } \lambda D(p_t)<c, \\
\in[0,1] & \text{if } \lambda D(p_t)=c,
\end{cases}
\end{align}
at all times $t$ with no immediate inspection, i.e. $\s(p_t)<\infty$.
Further, at any such reputation, the value of quality evolves according to 
\begin{align} \label{eq:IncentiveDiff}
 \frac{\de}{\de t} D(p_t) = (r+\l)D(p_t) + \s(p_t)D(p_t) - \s(p_t)\Brac{U(1,1)-U(0,0)} .
\end{align}
\end{proposition}
The formal proof of this and all following results are in Appendix \ref{App:main proofs}.
The terms on the right-hand side of \eqref{eq:best-response} are independent of the current state $\th$. The agent’s marginal benefit of effort is the same in the low and the high state.
Following \cite{board2013reputation}, we focus on equilibria where the agent chooses the same action in both states even when he is indifferent. 
To build intuition for the evolution of $D(p_t)$, note  that at  any reputation $p$ that triggers an immediate inspection ($\s(p)=\infty$),  we have $D(p)= U(1,1)-U(0,0).$
The evolution of the value of quality $D(p)$ in \eqref{eq:IncentiveDiff} is driven by the probability of future inspections and the spread in the possible post-inspection payoffs, $U(1,1)-U(0,0)$.
Whenever the current inspection rate is 0, $D(p_t)$ increases exponentially at rate $r+\l$ as we move closer to the next inspection. In particular, if there are no inspections between two reputation levels $p_0$ and $p_1$, we have
\begin{align*}
 D(p_0) = e^{-(r+\l)\tau(p_0,p_1)}D(p_1).
\end{align*}
where $\tau(p_0,p_1)$ is the time for reputation to travel from $p_0$ to $p_1$ according to \eqref{eq:ReputationODE}.

While condition \eqref{eq:best-response} is necessary for optimality only at reputations that are not immediately inspected, throughout the paper we focus on equilibria in which the agent's strategy satisfies Equation \eqref{eq:best-response} at at all $p$ which are consistent with believed effort strategy $\tilde \eta $.
 This restriction
 rules out behavior that can be sustained in continuous time but would not be optimal in any discrete-time approximation: if the principal inspects the agent with certainty at some reputation $p$, then the reputation jumps immediately to 0 or 1. Hence, the effort  at $p$ has no effect the outcome and any level would be optimal. For the principal, however, the agent's anticipated effort at $p$ is crucial for her incentive to inspect. 
This introduces an indeterminacy where the principal can be driven to inspect with certainty at a reputation she would not inspect if the agent was believed to exert effort; while the agent is induced to drop the effort level to 0 at this isolated point.
The robustness requirement to impose \eqref{eq:best-response} at all reputations consistent with $\tilde \eta $ filters this out and ensures that the agent's strategy is also optimal against a small perturbation to the principal's strategy that lowers the inspection probability below 1.

\subsection{The principal's problem: approval and inspections}
\label{subsec:principal}

\paragraph{Optimal approval policy.}
To determine optimal principal-strategies, fix believed and actual effort strategy $\tilde \eta = \eta$. 
Since these are functions of the reputation only, and the reputation is not affected by approval, it intuitive and easy to see from \eqref{eq:V recursive} that 
 $\alpha =1$ is optimal for the principal if and only if the expected flow payoff, $v(p) = pH - (1-p)L$, is positive. That is, whenever the reputation is high enough:
\begin{proposition}\label{lm:approval decisions}
The principal's approval strategy is optimal if and only if, at all times $t$, she approves ($\a_t =1$) if $p_t > p^\dagger:=L/(H+L)$ and does not approve  ($\a_t =0$) if $p_t<p^\dagger$. 
\end{proposition}
\noindent
We say the principal \textit{trusts} the agent when she grants approval and \textit{distrusts} when she does not grant approval. Proposition~\ref{lm:approval decisions} shows that the principal's approval is myopic. The inability to punish the agent when it is not optimal is familiar in foreign aid relations and known as ``Samaritan Dilemma'' \citep[][p. 340]{bourguignon2020foreign}. 

\paragraph{Optimal inspection policy.} We incorporate the optimal approval strategy from above and write the principal's flow payoff as $\max_\a \{ \a v(p) \} = v(p)^+$, where $v^+:=\max\{v,0\}$.  Under an optimal policy, the principal's value function $V$ satisfies the HJB equation\footnote{Writing the first expression with derivative $V'$ uses that the value function $V$ will be differentiable. A more permissive version of the quasi-variational inequalities that simplify to \eqref{eq: Principal HJB general} under differentiability are given in Appendix \ref{Apx:fixed point}. The formal proof of Theorem~\ref{thm:EQ characterization} in Appendix \ref{App:main proofs} provides closed-form expressions of $V$ for all equilibrium cases and confirms that $V$ is differentiable.}
\begin{multline}\label{eq: Principal HJB general}
 0 \ =\ \max \Big\{  v(p_t)^+ + V'(p_t) \dot p_t - r V(p_t)\  ; 
 \ (1-p_t)V(0)+ p_t V(1) - k - V(p_t) \Big\}.
\end{multline}
 Consider the first term, corresponding to continuing without inspection. Inserting $\dot p_t = \l\Paren{\tilde \eta(p_t) - p_t}$ the optimality condition for the no-inspection region can be rewritten as 
\begin{align}\label{eq:principal HJB waiting}
 r V(p_t) = v(p_t)^+ + V'(p_t) \l\Paren{\tilde \eta(p_t) - p_t}. 
\end{align}
This differential equation has the classic asset value interpretation: the continuation value in flow terms $r V(p_t) $ is equal to the sum of the current flow payoff $  v(p_t)^+ $ and the capital gains from changes in beliefs 
$
V'(p_t) \l\Paren{\tilde \eta(p_t) - p_t}.
$

Consider the second term in the maximization operator in \eqref{eq: Principal HJB general} corresponding to inspections. Denote by $$\Phi(p) = (1-p)V(0) + p V(1) - k $$ the expected net value for the principal from performing an inspection at current reputation $p$.
For reputations that are inspected, the HJB simply states
$V(p) = \Phi(p),$
i.e. the value function is equal to the value from inspecting. The boundary values $V(0)$ and $V(1)$ reached after inspection are determined endogenously within the fixed-point problem \eqref{eq: Principal HJB general}.

To determine which reputations are inspected,  
 suppose inspection is optimal at some level $p$. Then, the potential gain from waiting, given by the first term in \eqref{eq: Principal HJB general}, cannot exceed 0. With the value matching condition $V(p) = \Phi(p)$, inspection is optimal at reputation $p$ only if 
\begin{align}\label{eq:Phi}
 v(p)^+ + \Phi'(p) \l\Paren{\tilde \eta(p)-p} - r \Phi(p) \le 0.
\end{align}
 The left-hand side of the inequality is increasing in $\tilde \eta(p)$ because $\Phi'(p) = V(1)-V(0)>0$. 
This shows that, all else equal, the principal is less likely to inspect when the agent exerts more effort at this instant. In this sense, the agent's effort is a contemporaneous strategic substitute for the principal's inspection incentives (see discussion of contemporaneous and intertemporal strategic effects in Section \ref{sec:Multiplicity}).
It will be convenient to solve \eqref{eq:Phi} with equality to get a function $ \bar \eta$ such that inspection is optimal only if $\tilde \eta(p) \le \bar \eta (p)$, and it is optimal to mix between inspecting and waiting at $p$ if and only if $\tilde \eta(p) = \bar \eta (p)$, where $\bar \eta$ is 
\begin{align} \label{eq:eta bar}
 \bar \eta(p) = \frac{r\Paren{V(0)-k}}{\l \Paren{V(1)-V(0)}} + p \frac{r+\l}{\l} - v(p)^+ \frac{1}{\l \Paren{V(1)-V(0)}}. 
\end{align}
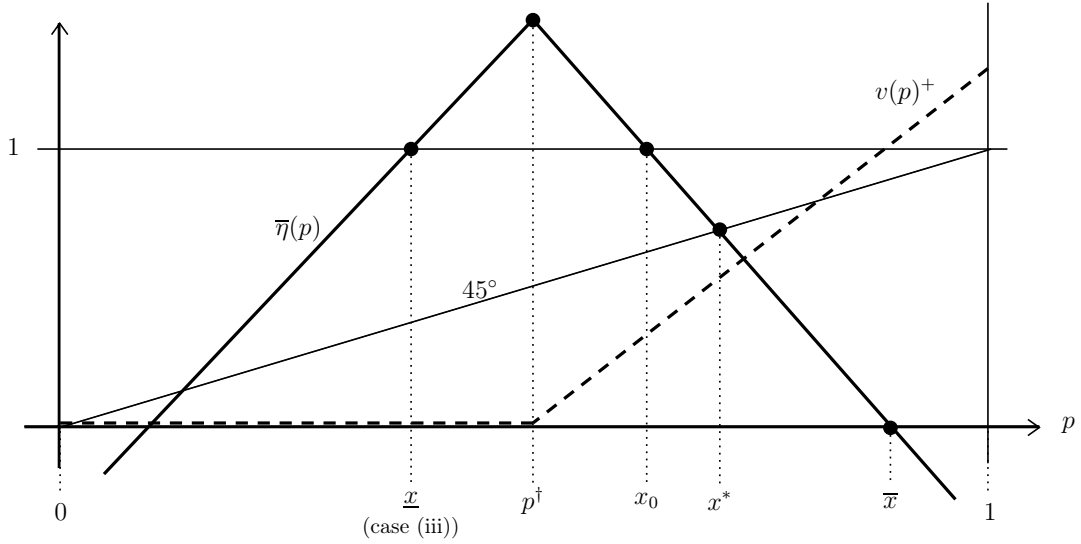
\begin{figure}[t!]\centering
\resizebox{.9\textwidth}{!}{
\begin{tikzpicture}[x=0.75pt,y=0.75pt,yscale=-1,xscale=1]

 \draw  [dash pattern={on 0.84pt off 2.51pt}]  (49,300) -- (49,340) ;
 \draw  (27,297) -- (651.62,297);
 \draw  (27,297) -- (651.62,297)(48,47) -- (48,322.97) (644.62,292.79) -- (651.62,297.79) -- (644.62,302.79) (43.74,54) -- (48.74,47) -- (53.74,54)  ;
 \draw    (35,125) -- (632,125) ;

\draw  [dash pattern={on 0.84pt off 2.51pt}]  (620,300) -- (620,340) ;

\draw [line width=1.5]  [dash pattern={on 5.63pt off 4.5pt}]  (49,295) -- (340,295) ;
\draw [line width=1.5]  [dash pattern={on 5.63pt off 4.5pt}]  (620,75) -- (340,295) ;
 \draw [fill={rgb, 255:red, 0; green, 31; blue, 0 }  ,fill opacity=1 ]   (48,298)  -- (622,125) ;
  \draw [color={rgb, 255:red, 0; green, 0; blue, 0}  ,draw opacity=1 ]   (620,34.19) --  (620,320) ;

\draw (180,165) node [anchor=north west][inner sep=0.75pt]  [font=\normalsize]  {$\overline{\eta }( p)$};
\draw (664.18,289.98) node [anchor=north west][inner sep=0.75pt]  [font=\normalsize]  {$p$};
\draw (616,343.4) node [anchor=north west][inner sep=0.75pt]  [font=\normalsize]  {$1$};

\draw (15,117) node [anchor=north west][inner sep=0.75pt]  [font=\normalsize]  {$1$};
\draw (549,80) node [anchor=north west][inner sep=0.75pt]  [font=\normalsize]  {$v( p)^{+}$};

\draw (295,205) node [anchor=north west][inner sep=0.75pt]  [font=\normalsize]  {$45^{\circ }$};
\draw (44,343.4) node [anchor=north west][inner sep=0.75pt]  [font=\normalsize]  {$0$};

 \draw (27,297.79) -- (651.62,297.79)(48.74,47) -- (48.74,322.97);

  \draw [line width=1.5] (76,326.69) -- (340,44.69) -- (600.01,342.39);
 \draw [dash pattern={on 0.84pt off 2.51pt}] (340,45) -- (340,333);
    \draw[fill=black] (340,45)circle (3pt);
  \draw (340,343) node {\normalsize $p^\dagger$};

  \draw[dash pattern={on 0.84pt off 2.51pt}] 
     (410,125) -- (410,333);
       \draw[fill=black] (410,125) circle (3pt);
\draw (410,343) node {\(\displaystyle x_0\)};

\draw[dash pattern={on 0.84pt off 2.51pt}] 
     (455,175) -- (455,333);
\draw (455,343) node {\(\displaystyle x^*\)};
  \draw[fill=black] (455,175) circle (3pt);

\draw[dash pattern={on 0.84pt off 2.51pt}] 
     (560,298) -- (560,333);
       \draw[fill=black]   (560,298) circle (3pt);
\draw (560,343) node {\(\displaystyle \overline{x}\)};

  \draw[dash pattern={on 0.84pt off 2.51pt}] 
     (265,125) -- (265,333);
       \draw[fill=black] (265,125) circle (3pt);
\draw (265,343) node {\(\displaystyle \underline x\)};
\draw (265,360) node {\footnotesize (case (iii))};

\end{tikzpicture}
}
\caption{The dashed line shows the principal's flow payoff under the optimal approval strategy. The bold line corresponds to the effort threshold $\bar \eta$ for inspection. If $\bar \eta(p) >1$, then the principal strictly prefers to inspect, regardless of the agent's effort.}\label{fig:eta}
\hfill
\end{figure}

The critical effort level $\bar \eta$, shown in Figure~\ref{fig:eta}, is piecewise linear and maximal at the indifference threshold $p^\dagger$. When the uncertainty about her optimal approval action is highest, information about the state is most valuable. It is never optimal for the principal to inspect at extreme reputations regardless of the agent's effort, and hence we have $\bar\eta(p)<0$ for $p$ sufficiently close to 0 or 1.

\subsection{Equilibrium characterization} \label{subsec:equilibrium charact}

The previous subsections describe the agent's best response to an inspection strategy for some boundary values $U(0,0)$ and $U(1,1)$; and the  principal's best response to an effort strategy for some boundary values $V(0)$ and $V(1)$.  To characterize  equilibria, we combine those observations and determine profiles of mutual best response together with the resulting boundary values.

In all that feature inspections, the space $[0,1]$ partitions into three regions: a \textit{blind trust} region $(\bar x, 1]$ at the top at which the principal approves without inspection, a \textit{blind distrust} region $[0, \underline x)$ at the bottom at which the principal neither approves nor inspects, and the \textit{inspection} region $[\underline x, \bar x]$ at which the principal inspects with positive rate or immediately. 
The following theorem collects the formal description of the three possible equilibrium types. We take as given the principal uses the optimal approval strategy with threshold $p^\dagger$ from Proposition~\ref{lm:approval decisions}. The theorem presents the three equilibrium classes in order of increasing effort. 

\begin{theorem}[Equilibrium characterization] \label{thm:EQ characterization}
There exist three cutoffs $ \underline x$, $x^*$, $\bar x $ satisfying $0< \underline x< p^\dagger < x^*<\bar x < 1$, such that at least one the following strategy combinations constitutes an equilibrium.
\begin{enumerate}[$(i)$]
 \item\label{thm:P1} The principal either inspects immediately iff $p \in [\underline x, \bar x]$, or she never inspects. The agent's effort is $\eta(p)=0$ for all $p\in [0,1]$. 
 \item\label{thm:P2} The principal inspects immediately if $p \in [\underline x, x^*)$, inspects at constant hazard rate $\s^* \in (0,\infty)$ if $p \in [x^*, \bar x]$, and does not inspect otherwise. The agent's effort is $\eta(p) = \frac{x^* (\bar x-p)}{\bar x - x^*}$ at reputations $p \in [x^*, \bar x)$, $\eta(p) =0$ otherwise. 
 
 \item\label{thm:P3} The principal inspects immediately if $p \in [\underline x,x_0 ] $, inspects at constant hazard rate $\s^*  \in (0,\infty)$ if $p \in (x_0, \bar x]$ and does not inspect otherwise, where $x_0=1+\bar x-\bar x/x^*$. The agent's effort strategy is 
 \[
 \eta(p) = \begin{cases}
 1 &\text{if } p \in [0, x_0), \\
 \frac{x^* (\bar x-p)}{\bar x - x^*} &\text{if } p\in [x_0 ,\bar x], \\
 0 &\text{if } p \in (\bar x, 1].
 \end{cases}
 \]
\end{enumerate}
Moreover, every equilibrium satisfying the robustness restriction is outcome equivalent to one of the above. 
\end{theorem}
We will focus the following discussion on the equilibrium class (iii), which exhibits the richest dynamics and which we call \textit{recovery equilibria}. It is the only class of equilibria in which the relationship persists indefinitely. 
 Classes (ii) is structurally identical to (iii) during the trust phase, but differs in that the relationship breaks down when $p=0$ is reached after the first failed inspection. We therefore refer to equilibria of class (ii) as \textit{breakdown equilibria}. The equilibrium in (i) is special in the sense that the agent exerts no effort, and the principal's problem is that of optimally inspecting an exogenous randomly declining state. If the principal inspects at all in this case, she inspects periodically. 

The structure of a recovery equilibrium (case (iii)) can be summarized as follows. Whenever an inspection reveals high quality, the principal grants an inspection holiday during which the agent exerts no effort and his reputation decreases gradually toward cutoff $\bar x$. Eventually, reputation reaches the cutoff $\bar x$, and then the principal starts to inspect the agent randomly at constant rate $\s^*$. Since $\bar x > p^\dagger$, this region displays ``trust but verify.'' 
The agent's effort is equal to $\bar \eta(p)$ which increases over time while the agent's reputation continues to decline. 
 The reputation never reaches the threshold $x^*$, however, as $\dot p_t$ approaches 0 as $p$ approaches $x^*$. 
Once the agent is eventually inspected, either the inspection reveals high quality and this cycle repeats, or the agent fails the inspection and the relationship moves into a recovery phase starting with $p=0$. In the recovery phase, the agent exerts maximum effort to regain the principal's trust. After a fixed time period with blind distrust, the principal performs a deterministic inspection once the reputation has increased to $\underline x$. If the inspection reveals low quality again, reputation resets to $p=0$ and the recovery cycle restarts. Once an inspection in the recovery phase reveals high quality, the agent's reputation is restored to $p=1$, and the game returns to blind trust as above. In this case (iii) note that reputations outside of $[0,\underline x] \cup (x^*, 1]$ are never reached on the equilibrium path. In particular, since $\underline x < p^\dagger < \bar x$, the principal only changes her approval decision after confirming by inspection that a change is optimal.

The equilibrium properties and multiplicity are intricately tied to the intertemporal strategic interactions at different levels of trust. We will discuss these and their implications for potential gains from communication after outlining the proof of Theorem~\ref{thm:EQ characterization}.

\subsection{Proof outline}

\begin{figure}
    \centering
    \resizebox{\textwidth}{!}{

\begin{tikzpicture}[x=0.75pt,y=0.75pt,yscale=-1,xscale=1]

\draw [color={rgb, 255:red, 74; green, 144; blue, 226 }  ,draw opacity=1 ][line width=5.25]  [dash pattern={on 5.9pt off 4pt}]  (336,190) -- (456,190) ;
\draw [color={rgb, 255:red, 74; green, 144; blue, 226 }  ,draw opacity=1 ][line width=5.25]    (229,190) -- (335,190) ;
\draw  [color={rgb, 255:red, 0; green, 0; blue, 0 }  ,draw opacity=1 ] (52.53,195.85) .. controls (52.52,200.52) and (54.85,202.85) .. (59.52,202.86) -- (129.26,202.92) .. controls (135.93,202.93) and (139.26,205.26) .. (139.25,209.93) .. controls (139.26,205.26) and (142.59,202.93) .. (149.26,202.94)(146.26,202.94) -- (218.99,203) .. controls (223.66,203.01) and (225.99,200.68) .. (226,196.01) ;
\draw [color={rgb, 255:red, 74; green, 144; blue, 226 }  ,draw opacity=1 ][line width=1.5]  [dash pattern={on 5.63pt off 4.5pt}]  (417.66,160.75) .. controls (480.91,94.68) and (534.17,113.2) .. (591.97,150.58) ;
\draw [shift={(594.61,152.3)}, rotate = 213.22] [fill={rgb, 255:red, 74; green, 144; blue, 226 }  ,fill opacity=1 ][line width=0.08]  [draw opacity=0] (8.75,-4.2) -- (0,0) -- (8.75,4.2) -- (5.81,0) -- cycle    ;
\draw [color={rgb, 255:red, 74; green, 144; blue, 226 }  ,draw opacity=1 ][line width=1.5]  [dash pattern={on 5.63pt off 4.5pt}]  (417.66,160.75) .. controls (375.64,-4.32) and (170.37,29.45) .. (58.68,144.26) ;
\draw [shift={(57,146)}, rotate = 313.77] [fill={rgb, 255:red, 74; green, 144; blue, 226 }  ,fill opacity=1 ][line width=0.08]  [draw opacity=0] (8.75,-4.2) -- (0,0) -- (8.75,4.2) -- (5.81,0) -- cycle    ;
\draw [color={rgb, 255:red, 0; green, 0; blue, 0 }  ,draw opacity=1 ]   (609.99,160.68) ;
\draw [shift={(609.99,160.68)}, rotate = 0] [color={rgb, 255:red, 0; green, 0; blue, 0 }  ,draw opacity=1 ][fill={rgb, 255:red, 0; green, 0; blue, 0 }  ,fill opacity=1 ][line width=0.75]      (0, 0) circle [x radius= 3.35, y radius= 3.35]   ;
\draw    (400.75,160.75) ;
\draw [color=black ,draw opacity=1 ]   (50.9,160) ;
\draw [shift={(50.9,160)}, rotate = 0] [color={rgb, 255:red, 0; green, 0; blue, 0 }  ,draw opacity=1 ][fill={rgb, 255:red, 0; green, 0; blue, 0 }  ,fill opacity=1 ][line width=0.75]      (0, 0) circle [x radius= 3.35, y radius= 3.35]   ;
\draw  [color={rgb, 255:red, 0; green, 0; blue, 0 }  ,draw opacity=1 ] (229.58,196.25) .. controls (229.58,200.92) and (231.91,203.25) .. (236.58,203.25) -- (345.64,203.25) .. controls (352.31,203.25) and (355.64,205.58) .. (355.64,210.25) .. controls (355.64,205.58) and (358.97,203.25) .. (365.64,203.25)(362.64,203.25) -- (444.1,203.25) .. controls (448.77,203.25) and (451.1,200.92) .. (451.1,196.25) ;
\draw [color={rgb, 255:red, 74; green, 144; blue, 226 }  ,draw opacity=1 ][line width=1.5]    (229,157) .. controls (281.02,85.33) and (446.34,-3.11) .. (603.63,148.69) ;
\draw [shift={(606,151)}, rotate = 223.51] [fill={rgb, 255:red, 74; green, 144; blue, 226 }  ,fill opacity=1 ][line width=0.08]  [draw opacity=0] (8.75,-4.2) -- (0,0) -- (8.75,4.2) -- (5.81,0) -- cycle    ;
\draw [color={rgb, 255:red, 74; green, 144; blue, 226 }  ,draw opacity=1 ][line width=1.5]    (229,157) .. controls (167.26,73.7) and (106.74,112.73) .. (61.74,150.68) ;
\draw [shift={(59,153)}, rotate = 319.52] [fill={rgb, 255:red, 74; green, 144; blue, 226 }  ,fill opacity=1 ][line width=0.08]  [draw opacity=0] (8.75,-4.2) -- (0,0) -- (8.75,4.2) -- (5.81,0) -- cycle    ;
\draw [color=black  ,postaction={decorate},
        decoration={
          markings,
          mark=at position 0.05 with {\arrow[scale=1.5]{<<}},
          mark=at position 0.2 with {\arrow[scale=1.5]{<<}},
          mark=at position 0.32 with {\arrow[scale=1.5]{<<}},
          mark=at position 0.5 with {\arrow[scale=1.5]{<<}},
          mark=at position 0.65 with {\arrow[scale=1.5]{<<}},
          mark=at position 0.8 with {\arrow[scale=1.5]{<<}},
          mark=at position 0.95 with {\arrow[scale=1.5]{<<}}
        }]   (366.44,160) -- (609.99,160) ;
\draw  [color={rgb, 255:red, 0; green, 0; blue, 0 }  ,draw opacity=1 ] (455.4,196.18) .. controls (455.47,200.85) and (457.83,203.15) .. (462.5,203.08) -- (520.82,202.31) .. controls (527.49,202.22) and (530.85,204.51) .. (530.91,209.18) .. controls (530.85,204.51) and (534.15,202.13) .. (540.82,202.04)(537.82,202.08) -- (599.14,201.27) .. controls (603.81,201.21) and (606.11,198.85) .. (606.05,194.18) ;
\draw [color=black  ,postaction={decorate},
        decoration={
          markings,
          mark=at position 0.05 with {\arrow[scale=1.5]{>>}},
          mark=at position 0.2 with {\arrow[scale=1.5]{>>}},
          mark=at position 0.35 with {\arrow[scale=1.5]{>>}},
          mark=at position 0.5 with {\arrow[scale=1.5]{>>}},
        }]   (50.9,160) -- (366.44,160) ;
\draw [line width=1.5]    (229,157) -- (229,168) ;
\draw [line width=1.5]    (293,157) -- (293,168) ;
\draw [line width=1.5]    (335.06,157) -- (335.06,168) ;
\draw [line width=1.5]    (366.44,157) -- (366.44,168) ;
\draw [line width=1.5]    (453.44,157) -- (453.44,168) ;
\draw  [color={rgb, 255:red, 74; green, 144; blue, 226 }  ,draw opacity=1 ] (228,231) .. controls (227.95,235.67) and (230.26,238.02) .. (234.93,238.07) -- (274.48,238.45) .. controls (281.15,238.52) and (284.46,240.88) .. (284.41,245.55) .. controls (284.46,240.88) and (287.81,238.58) .. (294.48,238.65)(291.48,238.62) -- (322.93,238.93) .. controls (327.6,238.98) and (329.95,236.67) .. (330,232) ;
\draw  [color={rgb, 255:red, 74; green, 144; blue, 226 }  ,draw opacity=1 ] (339,232) .. controls (339,236.67) and (341.33,239) .. (346,239) -- (393.11,239) .. controls (399.78,239) and (403.11,241.33) .. (403.11,246) .. controls (403.11,241.33) and (406.44,239) .. (413.11,239)(410.11,239) -- (447,239) .. controls (451.67,239) and (454,236.67) .. (454,232) ;

\draw (609.24,178) node  [font=\footnotesize]  {$p=1$};
\draw (142.65,220) node  [font=\footnotesize,color={rgb, 255:red, 0; green, 0; blue, 0 }  ,opacity=1 ] [align=left] {blind distrust};

\draw (534.18,220) node  [font=\footnotesize,color={rgb, 255:red, 0; green, 0; blue, 0 }  ,opacity=1 ] [align=left] {blind trust};

\draw (454.59,178) node  [font=\footnotesize]  {$\overline{x}$};

\draw (231,178) node  [font=\footnotesize]  {$\underline{x}$};

\draw (53.08,178) node  [font=\footnotesize]  {$p=0$};

\draw (367,178) node  [font=\footnotesize]  {$x^{*}$};

\draw (295.65,178) node  [font=\footnotesize]  {$p^{\dagger }$};

\draw (335.61,178) node  [font=\footnotesize]  {$x_0$};

\draw (352.88,220) node  [font=\footnotesize,color={rgb, 255:red, 0; green, 0; blue, 0 }  ,opacity=1 ] [align=left] {inspection};

\draw (285.88,257.19) node  [font=\footnotesize,color={rgb, 255:red, 0; green, 0; blue, 0 }  ,opacity=1 ] [align=left] {immediate};

\draw (405.88,257.19) node  [font=\footnotesize,color={rgb, 255:red, 0; green, 0; blue, 0 }  ,opacity=1 ] [align=left] {random};

\end{tikzpicture}
    }
  \caption{The figure shows the paths of trust in a recovery equilibrium. The interval shows the space of reputations from 0 to 1. Black arrows indicate the  reputation drift in the absence of inspections. The solid blue arcs represent an immediate inspection. Dashed blue arcs represent a random inspection. The equilibrium path includes all reputations in $[0,\underline x] \cup (x^*,1]$.}\label{fig:RecoveryEq}
\end{figure}
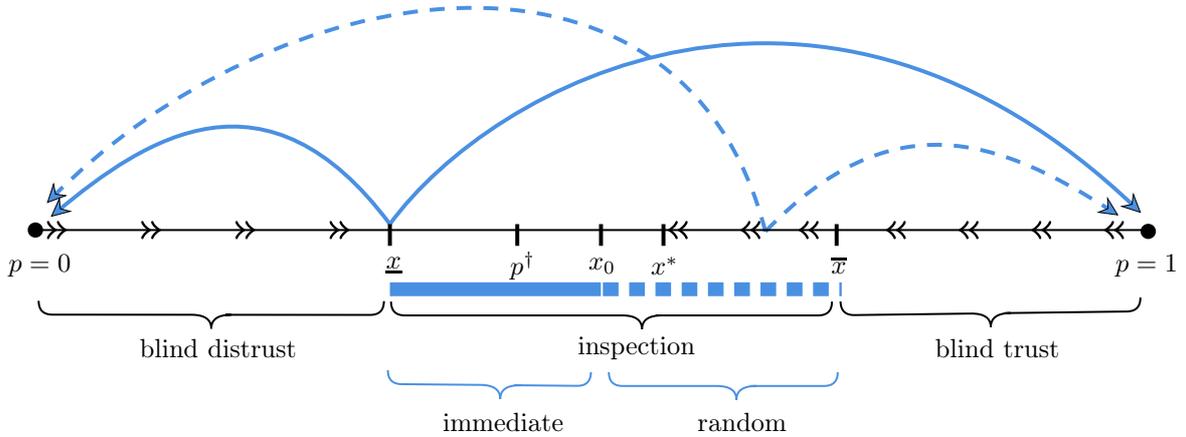

We examine each of the three reputation regions in isolation, starting with blind trust and blind distrust, and concluding with the inspection region. We focus on recovery equilibria presented in part (iii) of Theorem~\ref{thm:EQ characterization}. This class encompasses the full range of equilibrium behavior and it is straightforward to adapt the arguments to the other classes.

\paragraph{Blind trust.} Consider the region $(\bar x, 1]$. Intuitively, there must be a non-empty blind trust region in which the principal approves without inspecting, i.e. $\bar x <1$, because the principal faces a lump-sum cost of inspection, whereas the informational value of inspections becomes negligible as the reputation approaches 1. 
Since no inspections occur during blind trust, the agent's reputation must continuously decrease and reach $\bar x$ in finite time.
This implies that the agent's effort must be zero in the entire blind trust region. To see why, suppose to the contrary that $\eta(p_0)>0$ for some $p_0>\bar x$. From the agent's best-response condition \eqref{fig:eta}, it is optimal to exert positive effort only if $D(p_0) \ge c/\lambda$. From the evolution of $D(p_t)$ in Section \ref{subsec:agent} we know that for every $p\in (\bar x,p_0)$, the value of quality can be written as
\[
 D(p_0) = e^{-(r+\lambda)\tau(p_0,p)}D(p),
\]
where $\tau(p_0,p)>0$ is the time the agent's reputation takes to decrease from $p_0$ to $p$. Since $D(p_0)\ge c/\lambda$, this implies that $D(p)>c/\lambda$, so that the unique best response for the agent is to exert maximum effort at $p\in (\bar x,p_0)$. 
Then the reputation would increase: $\dot p=\lambda (1-p)>0$ for every $p\in (\bar x,p_0)$. This contradicts the fact that reputation must reach $\bar x$ in finite time. 
Since the agent must exert zero effort, inserting $ \tilde \eta(p_t) =0$ into \eqref{eq:principal HJB waiting} shows that the principal's value function while waiting satisfies the following ODE:
\begin{align}\label{eq:no effort HJB ODE}
rV(p)=v(p) -V'(p)\lambda p,   \qquad  \text{ for } p \in (\bar x,1].
\end{align}
The boundary $\bar x$ is depicted in Figure~\ref{fig:eta} as the highest reputation at which $\bar\eta(p) = 0$. Both the level $\bar x$ and the value $V(\bar x)$ are determined below when discussing the inspection region. With these, \eqref{eq:no effort HJB ODE} will give the boundary value $V(1)$.

\paragraph{Blind distrust.} Consider the region $[0,\underline x)$. Similar to the previous case, it is never optimal for the principal to inspect at reputations too close to certainty, so we must have $\underline x >0$. 
In a recovery equilibrium, the agent's effort must be maximal in the entire region $[0,\underline x]$. 
The reputation increases at 0 only if the agent is willing to exert effort at reputations around 0. 
Since effort incentives increase with proximity to the next inspection, similar to before, weak incentives at $p=0$, i.e. $D(0) \ge c/\l$ imply $D(p) > c/\l$ for all $p \in (0,\underline x]$. 
Since the agent exerts maximal effort, inserting $ \tilde \eta(p_t) =1$ into \eqref{eq:principal HJB waiting}, the principal's value function while waiting must satisfy the following ODE:
\begin{align}\label{eq:Recovery HJB unten}
rV(p)= V'(p)\lambda(1- p), \qquad    \text{ for } p \in [0,\underline x). 
\end{align}
The boundary $\underline x$ in a recovery equilibrium is depicted in Figure~\ref{fig:eta} as the reputation at which $\bar\eta(p) $ crosses 1 from below.

The level $\underline x$ and  value $V(\underline x)$ are determined below when discussing the inspection region. With these, \eqref{eq:Recovery HJB unten} will give the boundary value $V(0)$.

\paragraph{Inspection region.} Consider the region $[\underline x, \bar x]$.
For inspection to be optimal, the value function must satisfy 
\begin{align}\label{eq:HJB inspection region}
V(p)= \Phi(p), \qquad    \text{ for } p \in [\underline x, \bar x]. 
\end{align}
Here, the value from inspecting, $\Phi(p) = (1-p)V(0) + p V(1) -k$ depends on the continuation values after inspection, $V(0)$ and $V(1)$, which in turn
 depend on the reputation cutoffs $\underline x $ and $\bar x$ and their values. 
Continuation values and cutoffs are determined jointly via two value-matching conditions, $V(\underline x) = \Phi(\underline x)$ and $V(\bar x) = \Phi(\bar x)$, and two smooth-pasting conditions, 
$V'(\underline x) = V(1)-V(0) = V'(\bar x)$. 
Knowing that the agent exerts full effort on $[0,\underline x]$ and no effort on $(\bar x,1]$, we can fully identify the cutoffs and the principal's value. 
However, the principal's inspection behavior inside $[\underline x, \bar x]$ must induce the  agent to optimally choose those effort levels.

Consider first the upper part of the inspection region $[x^*, \bar x]$.
Since the agent must exert zero effort on $(\bar x, 1]$, the principal must inspect randomly at $\bar x$.
For the principal, it is optimal to inspect at random only if the agent exerts effort equal to the critical threshold $\bar \eta(p)$ in \eqref{eq:eta bar}. 
For the agent in turn, interior effort is optimal only if the value for quality $D(p_t) = c/\lambda$ continuously. Substituting this identity and $\frac{\de}{\de t}D(p_t)=0$ into Equation \eqref{eq:IncentiveDiff}, we can solve for the rate of inspection that makes it optimal to exert interior effort:
\begin{align}\label{eq:sigma*}
 \s^* = \frac{(r+\l)\frac{c}{\l}}{U(1,1)-U(0,0)-\frac{c}{\l}}.
\end{align}
Note that the equilibrium inspection rate is independent of the agent's reputation. 
The right-hand side of \eqref{eq:sigma*} implicitly depends on the inspection rate through its effect on the agent's continuation values $U(1,1)$ and $U(0,0)$. Thus, $\s^*$ is the solution to a fixed point problem which we show in the proof to exist uniquely.

At the lower boundary of the inspection region $\underline x$, an inspection occurs with probability one so that $D(\underline x)= U(1,1)-U(0,0)$. Reputations in $(\underline x,x^*]$ lie off the equilibrium path. 
The characterization in Theorem~\ref{thm:EQ characterization} also describes behavior for these reputations: For $ p \in (\underline x, x_0)$, with $x_0=1+\bar x-\bar x/x^*$ we have $\bar \eta(p) > 1$ so that the principal strictly prefers to inspect at any effort level. Consequentially, the agent-optimal effort level is 1 also on that interval. Starting from reputation $x_0$, which is the point at which $\bar \eta$ crosses 1 from above in Figure~\ref{fig:eta}, the agent chooses interior effort $\eta(p)$ and the principal inspects at constant rate $\s^*$ while the reputation drifts up toward $x^*$.

\begin{remark}[Lack off smooth pasting at $\underline x$ in cases (i) and (ii)]
    In contrast to the recovery equilibrium (iii), the lower boundary $\underline x$  in cases (i) and (ii) is determined differently: since the agent exerts no effort at low reputations, the inspection region expands and $\bar x$ is the reputation at which $\bar \eta$ crosses $0$ from below.
    In these cases the value function is equal to 0 on $[0,\underline x]$, which  implies that $0=V'(\underline x-) < \Phi'(p) = V(1)-V(0)$. This violation of smooth pasting is consistent with optimality in cases (i) and (ii) because within the lower waiting region $[0,\underline x)$ the reputation moves away from $\underline x$. See Appendix \ref{Apx:fixed point} for a formal derivation of the smooth-pasting conditions derived from quasi-variational inequalities. 
    Reputations in $(0,x^*]$ are off the equilibrium path in these cases.
\end{remark}

We observe that the recovery equilibrium features random inspections during trust but deterministic inspections during distrust. 
Further, the characterization in Theorem~\ref{thm:EQ characterization} shows that any equilibrium with effort has the same structure in the trust region but may vary in the distrust region. 
Classes (ii) and (iii) describe two different scenarios for equilibrium play during distrust: recovery and breakdown. 
The next section describes why equilibrium behavior is uniquely pinned down and involves mixing in the trust region, while there is multiplicity and pure strategies during distrust.

\subsection{Equilibrium~multiplicity~and~strategic~effects}

When the reputation lies in the trust region, the agent's and principal's strategic effects regarding effort and oversight work in opposite directions. The agent wants to exert effort only if he is inspected soon enough, while the principal wants to inspect only if the agent has not exerted too much effort. 
If effort is ever induced, this leads to a unique equilibrium structure at the top, with both parties using completely mixed strategies, determined by indifference conditions. The logic is similar to that of the static inspection game (for an example, see \citet[][p.19]{fudenberg1991game} as well as \cite{avenhaus2002inspection} and \cite{dresher1962} for repeated play of the period-by-period inspection problem).
Conversely, when the agent's reputation lies in the distrust region, the principal and agent's strategic effects go in the same direction. The agent wants to work if the principal inspects soon enough, and the principal is willing to inspect if the agent has exerted enough effort. 
This may lead to equilibrium multiplicity with players using deterministic, extremal strategies. The logic is akin to a coordination game.

The comparison to static games is intuitive, but incomplete because it does not account for the dynamic nature of the interaction and the persistent state . 
To better capture the nuances of the strategic interaction between the principal and the agent, we build on \cite{jun2004strategic} to extend the notion of strategic substitutability and strategic complementarity to our dynamic context. 

We distinguish contemporaneous and intertemporal strategic effects. For each player at a given time~$t$, we call the opponent's action a \textit{contemporaneous} strategic substitute (complement) if an anticipated increase in the opponent's current action decreases (increases) the player's incentive for their current action. We  saw in Section \ref{subsec:principal} that effort is a contemporaneous strategic substitute for inspections. 

Turning to dynamics, we call the opponent's action an \textit{intertemporal} strategic substitute (complement) if an increase in the opponent's past or future actions decreases (increases) the player's incentive for their current action. 
Note that the persistence in the state has different implications for the principal and the agent. The agent looks to the future because his effort affects the evolution of the state which will determine the outcome of future inspections. The principal looks to the past because the value of an inspection is influenced by the agent's past efforts. 

What is the intertemporal strategic effect of an increase in oversight on the agent's effort incentives? Here, an increase in oversight 
means that the next inspection will arrive sooner. It is easy to see that an anticipated increase in inspection intensity always boosts the agent's effort incentives, as a higher inspection intensity implies a higher value of quality $D(p)$. This means that oversight is an intertemporal strategic complement to the agent's effort, independent of the current level of trust. 

What is the intertemporal strategic effect of an increase in the agent's effort on the principal's inspection incentives? 
Here, the effect of past effort is aggregated through the reputation; the higher the expected accumulated past effort, the higher the current reputation.
To determine the intertemporal strategic effect of an increase in past effort, we consider the effect of a marginal increase in $p$ on the principal's benefit from waiting supposing that he is currently considering to inspect. 
Consider the principal's HJB in \eqref{eq: Principal HJB general}, suppose the value function at $p$ is $\Phi(p)$. We find that the marginal value for the principal from waiting without inspection is
\[
v(p)^++\Phi'(p)\dot p-r \Phi(p).
\]
Substituting the inspection value $\Phi(p)=pV(1)+(1-p)V(0)-k$ and reputational drift $\dot p=\lambda(\eta-p)$, yields
\begin{align}\label{eq:marginal value}
 v(p)^+ +(V(1) - V(0)) \lambda (\eta - p) - r (p V(1) + (1 - p)V(0)) - r k.
\end{align}
The marginal effect of a change in the agent's reputation is obtained by taking the derivative with respect to $p$:
\[
\alpha(p) (H +L) -(r+\lambda)(V(1) - V(0)).
\]
Note that we ignore the effect of a change in $p$ on $\eta$, as any effect arising from a change in the agent's current action at time $t$ is a contemporaneous effect, rather than an intertemporal one. In the proof of Theorem~\ref{thm:EQ characterization} we show that any equilibrium with inspections satisfies 
$
H +L >(r+\lambda)(V(1) - V(0)).
$
It therefore follows that in the trust region, where $\alpha(p)=1$, the marginal effect of an increase in reputation is positive. Waiting becomes more attractive, implying a decrease in the principal's incentive to inspect. On the other hand, in the distrust region, where $\alpha(p)=0$, the marginal effect of an increase in reputation is negative. Waiting becomes less attractive, implying an increase in the principal's incentive to inspect. In other words, the agent's effort is an intertemporal substitute for the principal in the trust region, but an intertemporal complement in the distrust region.

In summary, oversight is an intertemporal complement to the agent's effort at all times, while effort is an intertemporal substitute for the principal's oversight during periods of trust and an intertemporal complement during periods of distrust. 
Hence, the principal and agent's intertemporal effects work in opposite directions in the trust region, but in the same direction in the distrust region. This explains why, if effort is ever induced, we have a uniquely determined behavior during trust, but multiple equilibrium outcomes during distrust. We will see in Section \ref{sec:Self-disclosure} how communication can mitigate the coordination problem during distrust. 
While the strategic effects outlined above explain the  equilibrium structure \textit{within} an inspection cycle, we caution that the above derivations are based on fixed boundary values, which give the continuation utility \textit{after} an inspection. Across inspection cycles, the effect of increased oversight  on the agent's effort incentives may well be detrimental as we show in the next section.

\section{The cost of cheap oversight}\label{sec:Multiplicity}

We now turn to the existence of different types of equilibria and explicitly take into account the effect of changes in parameters on continuation values after an inspection. 
We establish necessary and sufficient conditions for existence in terms of two variables, the inspection cost $k$ and the agent's approval value $u$. The following theorem summarizes our
findings.
 \begin{theorem}\label{thm:multiplicity of EQ} 
 Fix any feasible combination of parameters\footnote{I.e., satisfying the assumptions on parameters laid out in Section \ref{sec:Model}} $\Paren{H,L,c,\l,r}$.
 There exist bounds 
 $ k_B$ and $k_R$ such that 
 \begin{enumerate}[$(i)$]
 \item For all $k\leq k_B$, there exist $\bar u_P^k>\underline u_B^k>0$, decreasing in $k$, such that an equilibrium with periodic inspections exists for all $u\le \bar u_P^k$, and a breakdown equilibrium exists for all $u\ge \underline u_B^k$. 
 \item For all $k\leq k_R$, there is $\underline u_R^k>0$ such that a recovery equilibrium exists for all $u\ge \underline u_R^k$. Moreover, for $k\le \min\{k_B, k_R\}$, we have $\underline u_R^k > \underline u_B^k $.
 
 \end{enumerate}
 For all values $(k,u)$ outside of these parameter ranges, there is a unique equilibrium outcome without inspections and zero effort.
\end{theorem}

The theorem shows that the agent tends to exert more effort with a smaller cost of oversight $k$ and larger approval value $u$. 
We also observe that the equilibrium is generally not unique, implying that the relationship may benefit from coordination. 

 An increase in oversight raises the agent's effort incentives at any given reputation for fixed boundary values, but it also shortens the approval phase, which decreases $U(1,1)$ and may lower his incentives for effort globally, so across inspection cycles. Similarly, at high reputations an increase in the agent's effort improves his reputation and reduces the principal's incentive to inspect. As a result, effort and intensive oversight can mutually offset each other and this effect can give rise to alternative equilibrium configurations for certain ranges of parameters. For example, for $k\leq k_B$ and $u\in [\underline u_B,\bar u_P]$, there exist two equilibria: a breakdown equilibrium involving effort and low-intensity oversight as well as an equilibrium without effort with either periodic oversight for or no oversight when $k$ larger than some $k_P$.

The fact that increased oversight can crowd out effort has unintuitive consequences for equilibrium existence as the inspection cost varies. Consider a reduction in $k$, for example as the result of an improvement to the inspection technology. In response to the reduced cost, the principal optimally inspects more frequently. The reduction in inspection cost has thus a positive direct effect on the principal's payoff, but the increased monitoring intensity also reduces the net present value for the agent which can destroy the agent's effort incentives altogether. 
\begin{corollary}\label{thm:detrimental effects of lower k}
A reduction in inspection cost can destroy effort incentives in equilibrium. 
\end{corollary}

Formally, this is a direct consequence of Part (i) of Theorem~\ref{thm:multiplicity of EQ}, which states that $\underline u_B$ and $\bar u_P$ are decreasing in $k$. Thus, for $k\leq k_B$ and an approval value $u$ just slightly above $u_B$, a reduction in $k$ to a level $\tilde k<k$ will result in new threshold $\tilde u_B$ increasing above the approval value $u$, making the initial breakdown equilibrium with effort no longer viable. 

Our observation that a reduction in inspection cost can undermine effort incentives contributes to a growing body of literature on the adverse consequences of improved information in principal-agent relationships \citep[see][]{cremer1995arm,dewatripont1999economics, prat2005wrong, ramos2019partnership, che2024predictive}. In our setting, the adverse effect of a reduction in the cost of inspection is rooted in the fact that the agent's value from the relationship is in part derived from periods of blind trust. Reducing the cost of inspecting increases the frequency of inspections and shortens the blind trust period. This lowers the agent's value from the relationship and, in turn, reduces his effort incentives. 

\section{The value of self-disclosure} \label{sec:Self-disclosure}
As highlighted above, the relationship can always suffer an irreversible breakdown after the first failed inspection even when a recovery equilibrium is feasible. This is due to the coordination problem arising during distrust. 
We now show how communication from the agent to the principal can be powerful to improve coordination during distrust and help to avoid unnecessary inspections. 
Assume for this section that the agent observes the true state at all times and that at each $t\ge 0$, the agent can send a cheap-talk message $m_t \in \{\emptyset, \hat H\}$ to the principal, where $m_t = \emptyset$ means the agent makes no report.

\begin{theorem}[Self-disclosure]\label{thm:self-disclosure}
 There exist $\bar k >0 $ and $\bar u>0$  such that for all  $k<\bar k$ and $u>\bar u$, there are cutoffs $\underline x\le x_0<x^*<\overline x$ such that there is an equilibrium with disclosure with the following reporting, inspection, and effort strategies. 
\begin{enumerate}[$(a)$]
 \item At $p=0$, as soon as a transition to $\th = H$ occurs, the agent truthfully  reports $\hat H$;  while the state remains $\th = L$, he misreports $\hat H$ with rate $\l \frac{1-x_0}{  x_0}$. For $p \in (0,\underline x)$, the agent reports $\hat H$ with probability 1 if $\th = H$, and reports $\hat H$ with probability $\frac{p}{1-p} \frac{1-x_0}{x_0}$ if $\th = L$. For $p\in [\underline x, 1]$, the agent sends no report to the principal. 
\item The principal inspects immediately if $p \in [\underline x,x_0] $, inspects at constant rate $\s^*$ if $p \in (x_0, \bar x]$, and does not inspect otherwise.
\item The agent's effort strategy is 
 \[
 \eta(p) = \begin{cases}
 1 &\text{if } p \in [0, x_0), \\
 \frac{x^* (\bar x-p)}{\bar x - x^*} &\text{if } p\in [x_0 ,\bar x), \\
 0 &\text{if } p \in [\bar x, 1].
 \end{cases}
 \]
\end{enumerate}
\end{theorem}

\begin{figure}
    \centering

\resizebox{\textwidth}{!}{
\begin{tikzpicture}[x=0.75pt,y=0.75pt,yscale=-1,xscale=1]

\draw [color={rgb, 255:red, 74; green, 144; blue, 226 }  ,draw opacity=1 ][line width=5.25]  [dash pattern={on 5.91pt off 4.02pt}]  (336,190) -- (454,190) ;
\draw [color={rgb, 255:red, 74; green, 144; blue, 226 }  ,draw opacity=1 ][line width=5.25]    (230,190) -- (335,190) ;
\draw  [color={rgb, 255:red, 0; green, 0; blue, 0 }  ,draw opacity=1 ] (52.53,195.85) .. controls (52.52,200.52) and (54.85,202.85) .. (59.52,202.86) -- (129.26,202.92) .. controls (135.93,202.93) and (139.26,205.26) .. (139.25,209.93) .. controls (139.26,205.26) and (142.59,202.93) .. (149.26,202.94)(146.26,202.94) -- (218.99,203) .. controls (223.66,203.01) and (225.99,200.68) .. (226,196.01) ;
\draw [color={rgb, 255:red, 74; green, 144; blue, 226 }  ,draw opacity=1 ][line width=1.5]  [dash pattern={on 5.63pt off 4.5pt}]  (417.66,160.75) .. controls (480.91,94.68) and (534.17,113.2) .. (591.97,150.58) ;
\draw [shift={(594.61,152.3)}, rotate = 213.22] [fill={rgb, 255:red, 74; green, 144; blue, 226 }  ,fill opacity=1 ][line width=0.08]  [draw opacity=0] (8.75,-4.2) -- (0,0) -- (8.75,4.2) -- (5.81,0) -- cycle    ;
\draw [color={rgb, 255:red, 74; green, 144; blue, 226 }  ,draw opacity=1 ][line width=1.5]  [dash pattern={on 5.63pt off 4.5pt}]  (417.66,160.75) .. controls (375.64,-4.32) and (158.91,-39.97) .. (60.17,129.63) ;
\draw [shift={(58.69,132.2)}, rotate = 299.61] [fill={rgb, 255:red, 74; green, 144; blue, 226 }  ,fill opacity=1 ][line width=0.08]  [draw opacity=0] (8.75,-4.2) -- (0,0) -- (8.75,4.2) -- (5.81,0) -- cycle    ;
\draw [color={rgb, 255:red, 0; green, 0; blue, 0 }  ,draw opacity=1 ]   (609.99,160.68) ;
\draw [shift={(609.99,160.68)}, rotate = 0] [color={rgb, 255:red, 0; green, 0; blue, 0 }  ,draw opacity=1 ][fill={rgb, 255:red, 0; green, 0; blue, 0 }  ,fill opacity=1 ][line width=0.75]      (0, 0) circle [x radius= 3.35, y radius= 3.35]   ;
\draw    (400.75,160.75) ;

\draw [shift={(50.9,162.42)}, rotate = 0] [color=black  ,draw opacity=1 ][fill={rgb, 255:red, 0; green, 0; blue, 0 }  ,fill opacity=1 ][line width=0.75]      (0, 0) circle [x radius= 3.35, y radius= 3.35]   ;
\draw  [color={rgb, 255:red, 0; green, 0; blue, 0 }  ,draw opacity=1 ] (229.58,196.25) .. controls (229.58,200.92) and (231.91,203.25) .. (236.58,203.25) -- (345.64,203.25) .. controls (352.31,203.25) and (355.64,205.58) .. (355.64,210.25) .. controls (355.64,205.58) and (358.97,203.25) .. (365.64,203.25)(362.64,203.25) -- (444.1,203.25) .. controls (448.77,203.25) and (451.1,200.92) .. (451.1,196.25) ;
\draw [color={rgb, 255:red, 74; green, 144; blue, 226 }  ,draw opacity=1 ][line width=1.5]    (335,160.68) .. controls (371.08,89.01) and (527.68,14.03) .. (601.67,141.98) ;
\draw [shift={(602.78,143.93)}, rotate = 239.28] [fill={rgb, 255:red, 74; green, 144; blue, 226 }  ,fill opacity=1 ][line width=0.08]  [draw opacity=0] (8.75,-4.2) -- (0,0) -- (8.75,4.2) -- (5.81,0) -- cycle    ;
\draw [color={rgb, 255:red, 74; green, 144; blue, 226 }  ,draw opacity=1 ][line width=1.5]    (335,160.68) .. controls (284.35,18.95) and (172.24,28.26) .. (58.45,148.8) ;
\draw [shift={(56.73,150.63)}, rotate = 313.09] [fill={rgb, 255:red, 74; green, 144; blue, 226 }  ,fill opacity=1 ][line width=0.08]  [draw opacity=0] (8.75,-4.2) -- (0,0) -- (8.75,4.2) -- (5.81,0) -- cycle    ;
\draw [color=black  ,postaction={decorate},
        decoration={
          markings,
          mark=at position 0.05 with {\arrow[scale=1.5]{<<}},
          mark=at position 0.2 with {\arrow[scale=1.5]{<<}},
          mark=at position 0.33 with {\arrow[scale=1.5]{<<}},
          mark=at position 0.5 with {\arrow[scale=1.5]{<<}},
          mark=at position 0.65 with {\arrow[scale=1.5]{<<}},
          mark=at position 0.8 with {\arrow[scale=1.5]{<<}},
          mark=at position 0.95 with {\arrow[scale=1.5]{<<}}
        }]   (366.44,160.75) -- (609.99,160.68) ;
\draw  [color={rgb, 255:red, 0; green, 0; blue, 0 }  ,draw opacity=1 ] (455.4,196.18) .. controls (455.47,200.85) and (457.83,203.15) .. (462.5,203.08) -- (520.82,202.31) .. controls (527.49,202.22) and (530.85,204.51) .. (530.91,209.18) .. controls (530.85,204.51) and (534.15,202.13) .. (540.82,202.04)(537.82,202.08) -- (599.14,201.27) .. controls (603.81,201.21) and (606.11,198.85) .. (606.05,194.18) ;
\draw [color={rgb, 255:red, 245; green, 166; blue, 35 }  ,draw opacity=1 ][line width=1.5]    (70.13,153.98) .. controls (170.72,91.28) and (260,88.72) .. (330,153) ;
\draw [shift={(333,158)}, rotate = 237.31] [fill={rgb, 255:red, 245; green, 166; blue, 35 }  ,fill opacity=1 ][line width=0.08]  [draw opacity=0] (8.75,-4.2) -- (0,0) -- (8.75,4.2) -- (5.81,0) -- cycle    ;
\draw [color={rgb, 255:red, 245; green, 166; blue, 35 }  ,draw opacity=1 ][line width=1.5]    (39.65,167.98) .. controls (18.26,179.44) and (15.32,136.39) .. (44.81,152.3) ;
\draw [shift={(43.17,165.7)}, rotate = 143.06] [fill={rgb, 255:red, 245; green, 166; blue, 35 }  ,fill opacity=1 ][line width=0.08]  [draw opacity=0] (6.97,-3.35) -- (0,0) -- (6.97,3.35) -- cycle    ;
\draw 
        (50.9,162.42) -- (366.44,160.75) ;
\draw [line width=1.5]    (229,157) -- (229,168) ;
\draw [line width=1.5]    (293,157) -- (293,168) ;
\draw [line width=1.5]    (335,157) -- (335,168) ;
\draw [line width=1.5]    (366.44,157) -- (366.44,168) ;
\draw [line width=1.5]    (453.44,157) -- (453.44,168) ;

\draw (609.24,180.15) node  [font=\footnotesize]  {$p=1$};
\draw (142.65,221.38) node  [font=\footnotesize,color={rgb, 255:red, 0; green, 0; blue, 0 }  ,opacity=1 ] [align=left] {blind distrust};
\draw (534.18,221.38) node  [font=\footnotesize,color={rgb, 255:red, 0; green, 0; blue, 0 }  ,opacity=1 ] [align=left] {blind trust};
\draw (454.59,178.3) node  [font=\footnotesize]  {$\overline{x}$};
\draw (231,178) node  [font=\footnotesize]  {$\underline{x}$};
\draw (53.08,179.2) node  [font=\footnotesize]  {$p=0$};
\draw (367,179) node  [font=\footnotesize]  {$x^{*}$};
\draw (23.67,131.64) node  [font=\footnotesize,color={rgb, 255:red, 245; green, 166; blue, 35 }  ,opacity=1 ]  {$m = \emptyset$};
\draw (223.82,89.76) node  [font=\footnotesize,color={rgb, 255:red, 245; green, 166; blue, 35 }  ,opacity=1 ]  {$m = \hat H$};
\draw (295.65,177.97) node  [font=\footnotesize]  {$p^{\dagger }$};
\draw (352.88,220.11) node  [font=\footnotesize,color={rgb, 255:red, 0; green, 0; blue, 0 }  ,opacity=1 ] [align=left] {inspection};
\draw (335,178.44) node  [font=\footnotesize]  {$x_0$};
\end{tikzpicture}
}
\caption{Paths of trust in an equilibrium with disclosure. Orange arrows depict jumps after reports; blue arrows correspond to inspections. In contrast to the recovery equilibrium (see Figure \ref{fig:RecoveryEq}), reputation remains at zero until the agent reports an improvement, which moves his reputation to $x_0$, at which point the principal inspects.}
\label{fig:disclosure}
\end{figure}

The equilibrium dynamics are illustrated in Figure \ref{fig:disclosure}. The reputations on the equilibrium path are $\{0\}\cup \{x_0\} \cup (x^*,1]$. After a failed inspection, the agent exerts full effort but his reputation remains at zero until he reports $\hat H$. Upon observing report $\hat H$, the reputation jumps to $x_0$ and the principal inspects immediately. Thus, the agent has a strict incentive to report $\hat H$ when the state is $H$ and is indifferent when the state is $L$ since an untrue report of $\hat H$ is inspected immediately. 
To make it optimal for the principal to inspect after the agent reports $\hat H$, the agent must occasionally report $\hat H$ even when the state is still $L$. The rates and probabilities of such a false report are chosen such that the posterior reputation following a report is $x_0$.

Self-reporting shortens the inefficient lag until reputation is reset to 1 after a  transition and reduces the number of unnecessary inspections when the state has not transitioned. Informative communication is possible because this inspection is of mutual interest during distrust. The agent's report aids in coordinating the timing of inspections. 
During trust, interests are conflicting and self-disclosure cannot help in detecting a deterioration in quality earlier. Within the trust region, any report from the agent that would increase the reputation makes the principal less willing to inspect.  Because the agent's report is cheap and he dislikes being inspected while enjoying approval, he would always send those reports that minimize oversight. 

Finally, we show that recovery and self-disclosure are valuable for the principal whenever feasible. While this is intuitive, the equilibrium effects are ambiguous a priori: 
In an equilibrium with breakdown, the agent remains longer in blind trust than in an equilibrium with recovery. Thus, recovery increases the expected inspection cost during trust. Moreover, breakdowns pose a large threat to the agent, implying stronger effort incentives. Despite these countervailing effects, the principal always benefits from recovery. Making recovery more efficient through self-disclosure increases this benefit further. Let the principal's value functions in a breakdown, recovery, and self-disclosure equilibrium be $V_B$, $V_R$, and $V_D$, respectively. 
\begin{proposition}\label{prop:value comparison}
We have $V_B(p)\le V_R(p) \le V_D(p)$ for all $p\in [0,1]$. 
\end{proposition}
For the first inequality, we make use of the fact that the upper threshold $\bar x$ is higher in the recovery than in the breakdown equilibrium. This implies that the agent exerts more effort at all times in the recovery than in the breakdown equilibrium. Thus, if we combined the principal's strategy from the breakdown equilibrium with the agent's strategy from the recovery equilibrium, then the principal's payoff would be higher than it was in the breakdown equilibrium. Since the principal is playing a best response, her payoff in the recovery equilibrium must be even higher. For the disclosure equilibrium, the principal could simply ignore the information reported by the agent and would be at least as well off as in the recovery equilibrium. Since the additional information from the report only benefits her, she must be better off.

\section{Conclusion}\label{sec:conlusion}

We present a dynamic principal-agent model that examines the interplay between trust and oversight. We provide a comprehensive analysis of the equilibrium dynamics that may unfold in such relationships, and emphasize the importance of underlying contemporaneous and intertemporal strategic effects of effort and oversight incentives that shape the equilibrium structure. Our analysis reveals that oversight and  trust jointly undergo cycles. Reduced oversight costs can reduce efficiency as excessive oversight destroys effort incentives. We also find that communication between the principal and agent can play a vital role in resolving coordination problems in rebuilding trust. 

Our findings have implications for a wide range of real-world scenarios, such as foreign aid programs, supply-chain self-regulation, and government contracting. These insights can inform decision-makers on how to strike the right balance between trust and oversight to ensure the effectiveness, fairness, and success of various relationships and initiatives. In particular, simple communication procedures can reduce the inefficiencies associated with breakdowns in relationships, which are common in foreign aid as recipient countries cease reform efforts once aid flows are interrupted \citep[see][]{collier1997redesigning}.

An additional constraint in Foreign aid allocation that we do not consider in this paper is ``the competition among donors.''\footnote{With the ingredients of our model (i) moral hazard, (ii) costly monitoring, and (iii) the inability to commit or coordinate in MPE, our analysis captures three of the four major constraints faced by donors in Foreign aid relationships: (i) ``the ability of the recipient countries’ governments to deviate from donors’ intended
objectives,'' (ii)  ``the cost of effectively monitoring,'' and  (iii) ``the limited credibility of the threat [...] by donors if conditionality is not met -- i.e. the so called `Samaritan dilemma'' \citep[see][p. 308]{bourguignon2020foreign}. The fourth constraint is competition among multiple donors.} 
A problem that frequently arises in practice is that donors differ in their objectives or precise conditions they want to enforce. 
Our model can capture such a scenario if we interpret the agent's effort cost as also reflecting a cost from competing interests. 
While a full multi-donor model is beyond the scope of this paper, this would enable the analysis of other problems arising from multiple donors like free-rider or public-good provision problems.

\begin{singlespace} 
\bibliographystyle{chicago}
\bibliography{references}
\end{singlespace}
\newpage 
\setcounter{section}{0}
\begin{appendix}
\section{Proofs}\label{App:main proofs}

\subsection{Proof of Proposition~\ref{prop: effort incentives}}
To derive the effort incentives of the agent, consider the agent's problem in  \eqref{eq:U recursive}. Note that the evolution of the reputation is governed by the \textit{believed} effort strategy $\tilde \eta$, so that the derivative with respect to the agent's true effort $\eta$ is
$ \l U(p,1) - \l U(p,0) - c = \lambda D(p)-c.
$
Thus, the agent's effort strategy is optimal if and only if it satisfies the condition in the result statement for every instant $[t,t+\de t)$ without strictly positive inspection mass. 

To establish the time evolution of the value of quality, $D(p_t)$, we truncate the  the agent's continuation payoff at the first arrival of either a quality shock or an inspection, whichever comes first:
\begin{dense}
\begin{align}\label{eq:U truncated at insp}
U(p_t, \th) \ = \int_{t}^\infty e^{-(r+\l)(s-t)-\int_t^s \s(p_z)\de z } {\Big (} \alpha(p_s)u+ \eta(p_s)(\l D(p_s) -c) + \l U(p_s,0)  \ 
 + \ \s(p_s) U(\th, \th){\Big )}\de s.
\end{align}
\end{dense}
\noindent
To determine the evolution of $D(p_t)$, consider the right-hand side of expression \eqref{eq:U truncated at insp} at $\th=1$ and subtract the same expression for case $\th = 0$. Fixing the initial reputation $p$, this gives 
\begin{dense}
\begin{align*}
D(p_t) \ = \int_{t}^\infty e^{-(r+\l)(s-t)-\int_t^s \s(p_z)\de z } {\Big (}  \s(p_s) \big( U(1, 1) - U(0,0)\big) {\Big )}\de s.
\end{align*}
\end{dense}
 Taking the derivative with respect to $t$ gives the time derivative in the result statement for all reputations which are not immediately inspected.
\qed

\subsection{Proof of Proposition~\ref{lm:approval decisions}}
The principal's approval decision does not affect her belief $p_t$ about $\theta_t$, and therefore, it is optimal for the principal to give approval to the agent whenever $v(p_t)=p_t H-(1-p_t)L>0$ or equivalently $p_t>p^\dagger$ and to withdraw approval when $p_t<p^\dagger$. \qed

\subsection{Proof of Theorem~\ref{thm:EQ characterization}}\label{Apx:Thm1}
Our general approach is to use the HJB equations, together with the necessary value matching and smooth-pasting conditions, to derive closed-form expressions for the value functions. By construction, these candidate functions satisfy the HJB equations, which is a sufficient condition for optimality. The proof of Theorem~\ref{thm:EQ characterization} includes auxiliary results which are proved formally in Online Appendix~\ref{Apx:Aux}. 

It is easy to verify that the principal never inspects if $k$ is too large, in which case the agent clearly never exerts effort (as $D(p)=0$ everywhere). This equilibrium is included in part (i) of Theorem~\ref{thm:EQ characterization}. 
For the rest of the proof, assume that $k$ is low enough that inspection for the principal can be optimal. 

\paragraph{Equilibrium type (i): deterministic inspection at supremum.}
Suppose that an inspection occurs with probability one at $\bar x \equiv \sup\{p \colon \de N_{F}(p) >0 \}$, which will result in the no-effort equilibrium with inspection.  
Lemma \ref{lm:blind regions} and Lemma \ref{lm: immediate inspection at top implies no effort} in the online appendix confirm that immediate inspection at $\bar x$ implies that $D(\bar x)= U(1,1)-U(0,0) \le \frac{c}{\l}$ and that the equilibrium must feature \textit{breakdown} at reputation $p=0$. 
 That is, the agent never works again once an inspection revealed the low state. 
In this case, the only reputations on the equilibrium path are $\{0\} \cup [\bar x, 1]$, and the agent exerts no effort at any reputation other than (potentially) at exactly $\bar x$ if $U(1,1)-U(0,0) = \frac{c}{\l}$.\footnote{Note that the effort chosen at an isolated reputation $\bar x$ that is inspected immediately has no effect on the outcome path. 
Within type (i), we can therefore focus on equilibria in which the agent never exerts effort even if $U(1,1)-U(0,0) = \frac{c}{\l}$.}

 We confirm in any equilibrium with immediate inspection at $\bar x$, there must be an immediate inspection if and only if $p\in [\underline x, \bar x]$ and the cutoffs in this case satisfy $\underline x < p^\dagger < \bar x$.\footnote{Recall that, although $[0,\bar x)$ is off the equilibrium path in this case, we specify optimal behavior for these reputations as well.}

Consider the HJB for the principal from \eqref{eq: Principal HJB general} for the case that $\eta = 0$ always: 
\begin{align}\label{eq:Principal HJB in proof}
 0 = \max \left\{ v(p)^+ - \l V'(p) p - r V(p)   ;   \Phi(p)- V(p) \right\},
\end{align}
where $\Phi(p)=p V(1)+(1-p)V(0)-k$ denotes the value of inspection at reputation $p$. 
Given that $\eta = 0$ everywhere, the lower boundary condition is $V(0) = 0$. 
At the inspection cutoff $\bar x$, value matching and smooth pasting imply 
$$ \Phi(\bar x) - V(\bar x) = \bar x V(1)- k - V(\bar x) = 0 $$ 
and 
\begin{align} \label{eq:valuematching pbar}
\Paren{-L +\bar x (H+L)}^+ - \l\underbrace{ V(1)}_{=\Phi'(\bar x)} \bar x- r\underbrace{\Paren{\bar x V(1)- k }}_{=\Phi(\bar x)}= 0. 
\end{align} 
The smooth-pasting condition \eqref{eq:valuematching pbar} implies that $\bar x > p^\dagger$. Otherwise we would have 
$
\Paren{-L +\bar x (H+L)}^+ = 0
$, 
in which case \eqref{eq:valuematching pbar} would be equivalent to $r \Phi(\bar x) = - \l \bar x V(1)$. Since clearly $V(1)>0$, this would make inspection at $\bar x$ strictly suboptimal. Hence, we have $\bar x > p^\dagger$ and $\Phi(\bar x)>0$.

To see that $\underline x < p^\dagger$, note that for $p\le p^\dagger$, the first term on the right-hand side of the HJB equation \eqref{eq:Principal HJB in proof} is $0 - \l V'(p) p - r V(p)$. For no-inspection to be optimal, this term must be 0, which implies $V(p) =0$ constantly on any no-inspection range.
Thus, when the agent never exerts effort, for $p\le p^\dagger$, waiting and not inspecting can be optimal only if $\Phi(p) \le 0$ to satisfy \eqref{eq:Principal HJB in proof}.
Since $V(1) > 0$, $\Phi(p)$ is continuously increasing in $p$. Thus, there is an $\underline x$ satisfying $\Phi(\underline x) = 0$. The principal optimally inspects immediately at any $p \in [\underline x, p^\dagger]$. 

We confirm in Lemma \ref{lm:no gap in inspection region} that the principal also inspects immediately at any $p \in (p^\dagger, \bar x)$. 
Thus, for all $p \in (\underline x, \bar x)$, immediate inspection is optimal and the first term in the max operator on the right-hand side of \eqref{eq:Principal HJB in proof} ---corresponding to the no-inspection payoff--- remains strictly below 0.

Lastly, we check that $\underline x$ defined by $\Phi(\underline x) = 0$ satisfies $\underline x< p^\dagger $: for $p$ above $p^\dagger$  the function  $\Paren{-L +p(H+L)}^+ - p(r+\l) V(1) + r k$ is increasing in $p$, and it is 0 at $p= \bar x$ by \eqref{eq:valuematching pbar}.
 $\Paren{-L +p(H+L)}^+ - p(r+\l) V(1) + r k$ is strictly negative for $ p = p^\dagger - \e$ for $\e>0$ small enough, implying that $\Phi(p^\dagger -\e) > \frac{\l}{r+\l} k >0 $.
 Clearly, for $p<\underline x$, $\Phi(p)<0$ and the principal would never inspect again. 

We now confirm that the necessary conditions above indeed form an equilibrium. The ODE that characterizes the principal's HJB on $[\bar x, 1]$ and satisfies the value matching condition at (yet to be determined) reputation threshold $\bar x$ has the (candidate) solution 
\begin{align} \label{eq:value function free x bar}
 W(p) = - \frac{L}{r} + p \frac{H+L}{r+\l} + \Paren{\frac{\bar x}{p}}^{\frac{r}{\l}}\Brac{ \frac{L}{r} - \bar x \frac{H+L}{r+\l} -k + \bar x W(1)}.
\end{align}
Inserting $p= 1$ above delivers the value $W(1)$: 
\begin{align}\label{eq:W(1) in no effort eq}
 W(1) = - \frac{L}{r}\frac{1- \bar x^{\frac{r}{\l}}}{1- \bar x^{\frac{r+\l}{\l}}} + \frac{H+L}{r+\l} - k \frac{\bar x^{\frac{r}{\l}}}{1- \bar x^{\frac{r+\l}{\l}}}.
\end{align}
We can then use the smooth pasting condition $ \left. \frac{\partial W(p) }{\partial p} \right\vert_{p= \bar x} = W(1) $ to determine the cutoff $\bar x$. This results in the condition 
\begin{align*}
 - \frac{1}{\bar x} \frac{L - r k }{r+ \l}+ \frac{H+L}{r+\l} = - \frac{L}{r}\frac{1- \bar x^{\frac{r}{\l}}}{1- \bar x^{\frac{r+\l}{\l}}} + \frac{H+L}{r+\l} - k \frac{\bar x^{\frac{r}{\l}}}{1- \bar x^{\frac{r+\l}{\l}}}.
\end{align*}
Rearranging terms gives the equivalent condition 

\begin{align}\label{eq:SP rearranged}
 0 = \frac{1}{\bar x} \frac{1}{1- \bar x^{\frac{r+\l}{\l}}}\Brac{ \frac{L - r k }{r+ \l} - \bar x \frac{L}{r} + \bar x^{\frac{r+\l}{\l}} \frac{\l}{r} \frac{L- rk}{r+\l}}.
\end{align}
It is easily verified that the term in square brackets is decreasing in $\bar x$ for any $\bar x \in [0,1]$. As $\bar x \nearrow 1$, the value in square brackets is $-k$. Finally, at $\bar x = p^\dagger = \frac{L}{H+L}$, the term in square brackets is equal to 
\begin{align*}
 \frac{L}{H+L}\Paren{ \frac{H+L}{L} \frac{L - r k }{r+ \l} - \frac{L}{r} + \Paren{\frac{L}{H+L}}^{\frac{r}{\l}} \frac{\l}{r} \frac{L- rk}{r+\l}}.
\end{align*}
If $k$ is too large so that the term in parentheses above is negative, it is optimal for the principal to never inspect. Whenever $k$ is low enough, such that the term in parentheses is positive, continuity in $\bar x$ implies that there must be a cutoff $\bar x \in (p^\dagger,1)$ such that \eqref{eq:SP rearranged} is satisfied.

Given the value $\bar x$ determined above, we can compute the equilibrium value $V(1)$ with \eqref{eq:W(1) in no effort eq}. This, in turn gives the value for inspection payoff $\Phi(p) = p V(1) - k$ and $\underline x$ is then simply given as the reputation at which the inspection payoff is 0: 
$\underline x = k/V(1) $.
This characterization with immediate inspection at the supremum leads to the behavior described in part (i) of Theorem~\ref{thm:EQ characterization}. 

The next case with random inspection at the supremum will result in the equilibria described in parts (ii) and (iii) of Theorem~\ref{thm:EQ characterization}. 

\subsubsection*{Equilibrium types (ii) and (iii): random inspection at supremum.}

Suppose the principal mixes between inspecting and not inspecting at reputation $\bar x$. 
This random inspection has to arrive at a rate over time rather than with positive probability (an atom) at reputation $\bar x$ (see Lemma \ref{lm:random inspection at top implies rate}). 

To determine the equilibrium behavior for high reputations, in particular inspection rate $\s^*$ and the interior effort level in the region $\underline x, \bar x$, refer to the outline in the main text after Theorem~\ref{thm:EQ characterization} and the formal derivations in Lemma \ref{lm:random inspection at top implies mixed effort and constant inspection rate} and Lemma \ref{lm:fix point inspection rate} in Online Appendix \ref{Apx:Aux}.
Thus, the equilibrium behavior for high reputations must follow one of the three cases outlined in Theorem~\ref{thm:EQ characterization}, where the reputations that can be reached from above and are on the equilibrium path are $[\bar x, 1]$ in part (i) and $(x^*, 1]$ in parts (ii) and (iii). 

We next turn to low reputations that can be reached only after an inspection revealed the low state.
First, note that whenever the believed effort strategy is such that the agent exerts no effort when the state is low with certainty, i.e. $\tilde \eta(0)=0$, then the agent's reputation (absent inspection) remains at 0, and by Lemma~\ref{lm:blind regions} the principal never inspects again. 
For the agent in turn, $\eta(0)=0$ is optimal since he will never be inspected. In any such equilibrium, once an inspection reveals that the current state is low, the reputation will remain at 0 forever and both players become inactive forever onward. This describes the behavior at lower reputation in parts (i) and (ii) of Theorem. 

Note that the optimality conditions for the principal outlined in equations \eqref{eq:value function free x bar}-\eqref{eq:SP rearranged} for the equilibrium part (i) are identical for the equilibrium in part (ii) so that $\underline x$ and $\bar x$ are the same.\footnote{Lemma \ref{lm:no effort at top} confirms that there is no effort at reputations $p\in(\bar x, 1]$.} 
To see this, note that in both cases $V(0)=0$ for the principal and the value-matching and smooth-pasting conditions at $\bar x$ following \eqref{eq:value function free x bar} are equivalent to both arguments on the right-hand side of \eqref{eq:HJB proof part B} being 0 with $\bar \eta(\bar x) = 0$. Intuitively, since inspection and effort strategy are the same in both equilibria on $[\bar x, 1]$ and the principal is indifferent between inspection and waiting at $\bar x$ in case (ii), it follows that also $\bar x$ and $V(1)$ are the same in both equilibria. 
Note however that the agent's utility, in particular $U(1,1)$, will not be the same across these two equilibria because the principal chooses different inspection strategies.

\subsubsection*{Equilibrium type (iii): recovery.}

Lastly, we consider part (iii) of the Theorem in which the agent does exert effort after a failed inspection.
The equilibrium behavior at high reputations $p \in [x^*, 1]$ is the same as in part (ii).\footnote{However, with a different value for the cutoff $\bar x$.} We thus focus first on the lower reputations. 
Suppose that $\eta(0) >0$ is optimal for the agent. Then at $p_t=0$ we have $D(p_t) \ge c/\l$ . By Lemma~\ref{lm:blind regions} there is no inspection on a non-empty interval $[0,\underline x)$. It follows that $D(p_t)$ must be strictly increasing as long as there is no inspection (see Equation~\ref{eq:IncentiveDiff}). Thus, the agent must exert full effort on reputations $(0,\underline x)$.\footnote{Again the exact effort level at isolated reputation $p=0$ does not affect the outcome path as long as effort is strictly positive; we thus set $\eta(0)=1$ in this class without loss up to equivalence.}

The principal's value function on $[0,\underline x)$ is then determined by the ODE \eqref{eq:Recovery HJB unten} which we restate here:
\begin{align}\label{eq:Recovery HJB unten in proof}
rV(p)= V'(p)\lambda(1- p)    \text{ for } p \in [0,\underline x). 
\end{align}
The value matching condition at the lower boundary $\underline x$ is 
\[
\Phi(\underline x) = V(0) + \underline x\Paren{V(1)-V(0)} - k.
\]

Together with the characterization at high reputations in Equation \eqref{eq:value function free x bar}, the candidate solution $W(p)$ interpreted as a function of the yet to be determined thresholds $\underline x$ and $\bar x$ given by 
\begin{align}\label{eq:W(p)}
 W(p) \ = \ \begin{cases}
 \brac{\frac{\bar x}{p}}^{\frac{r}{\l}}\Brac{ \frac{L}{r} - \bar x \frac{H+L}{r+\l} + W(0) + \bar x \Delta W- k} 
 - \frac{L}{r} + p \frac{H+L}{r+\l} &\text{ if } p \ge \bar x, \\
 W(0) + p \Delta W- k &\text{ if } p \in [\underline x, \bar x], \\
 \Paren{\frac{1-\underline x}{1-p}}^{\frac{r}{\l}}\Brac{ W(0) + \underline x \Delta W- k} &\text{ if } p < \underline x.
 \end{cases}
\end{align}
where $\Delta W=W(1)-W(0)$. Next, insert $p=0$ and $p=1$ above to determine $W(0) $ and $W(1) $. 
Finally, use the smooth-pasting condition at $\underline x$, 
\[
\left.\frac{\partial W(p) }{\del p} \right\vert_{p \nearrow \underline x} = W(1) - W(0),\] 
and the smooth pasting condition at $\bar x$, 
\[
\left.\frac{\partial W(p)}{\del p} \right\vert_{p\searrow \bar x} = W(1) - W(0),
\]
to determine the cutoffs $\underline x$ and $\bar x$. After some algebra, the conditions determining these cutoffs are obtained as
\begin{dense}
\begin{align}\label{eq: xU condition}
 0 &\ = \ \frac{\left(\lambda (H+L) \bar{x}^{\frac{\lambda +r}{\lambda }}+H r-\lambda L\right)}{(\lambda +r)\left(\lambda \left(\bar{x}^{\frac{\lambda +r}{\lambda }}+(1-\underline{x})^{\frac{\l+r}{\lambda }}\right)+r\right)}\left(r \underline{x}+\lambda \left((1-\underline{x})^{\frac{\lambda +r}{\lambda } }-(1-\underline{x})\right)\right) -r k,
\end{align}
\end{dense}
\noindent and 
\begin{multline}
 0 = \ \frac{\left(\lambda (H+L) \bar{x}^{\frac{\lambda +r}{\lambda }}+H r-\lambda L\right)}{(\lambda +r) \left(\lambda \left(\bar{x}^{\frac{\lambda +r}{\lambda }}+(1-\underline{x})^{\frac{\lambda +r}{\lambda }}\right)+r\right)}\left(r \bar{x}+\lambda \left((1-\underline{x})^{\frac{\lambda +r}{\lambda }}+\bar{x}\right)\right) \\ -r k-H \bar{x}+L (1-\bar{x}). \label{eq: xO condition}
\end{multline}
First, observe that the term $\left(r \underline{x}+\lambda \left((1-\underline{x})^{\frac{\lambda +r}{\lambda } }-(1-\underline{x})\right)\right)$ in \eqref{eq: xU condition} is positive for all $\underline x \in[0,1]$. Hence, \eqref{eq: xU condition} can be satisfied, and recovery equilibria exist, only if the numerator in the quotient in the first line is positive. Note further that the first term in \eqref{eq: xO condition} is strictly larger than the first term in \eqref{eq: xU condition}. Moreover, the last term $-H \bar{x}+L (1-\bar{x}) $ in \eqref{eq: xU condition} is negative if and only if $\bar x > p^\dagger$. It follows that both equations can hold simultaneously only if $\bar x > p^\dagger$.

Next, consider the right-hand side of \eqref{eq: xU condition} at $\underline x =0 $. The first term becomes 0 and the right-hand side is thus $-rk$. This term is strictly increasing in $\underline x$ if and only if the numerator of the fraction is positive, as required previously. 
Note that we need $\underline x < p^\dagger$, as otherwise the principal would not be willing to inspect at any reputation at which the agent exerts full effort. As illustrated in Figure~\ref{fig:eta}, at reputation $p^\dagger$, the principal's incentive to inspect is the largest. 
Thus, if she were unwilling to inspect on $[0,p^\dagger]$ while the agent exerts full effort, she would not inspect at any reputation at which the agent exerts full effort. This would rule out recovery. 
Therefore, we consider the right-hand side of condition \eqref{eq: xU condition} at $\underline x \nearrow p^\dagger $. This is equal to 
\begin{dense}
\begin{align*}
 \frac{\left(\lambda (H+L) \bar{x}^{\frac{\lambda +r}{\lambda }}+H r-\lambda L\right) }{(\lambda +r)\left(\lambda \left(\bar{x}^{\frac{\lambda +r}{\lambda }}+(\frac{H}{L+H})^{\frac{\l+r}{\lambda }}\right)+r\right)}\left(r \frac{L}{L+H}+\lambda \left(\Brac{\frac{H}{L+H}}^{\frac{\lambda +r}{\lambda } }-\frac{H}{L+H}\right)\right)-r k. 
\end{align*}
\end{dense}
\noindent For $k$ small enough, this term is strictly positive so that by continuity for any $\bar x \in (p^\dagger, 1)$, there must be a cutoff $\underline x \in(0,p^\dagger)$ such that condition \eqref{eq: xU condition} holds. 

Finally, consider the right-hand side of \eqref{eq: xO condition} at $\bar x =1 $. The first term becomes 0 and the right-hand side is thus $-rk-H$.
The right-hand side of \eqref{eq: xO condition} is strictly decreasing in $\bar x$ and we have already established that it is positive at $\bar x \searrow p^\dagger$ because $-H \bar{x}+L (1-\bar{x}) $ is equal to $0$ at $\bar x = p^\dagger$. Thus, by continuity there is a $\bar x \in (p^\dagger , 1)$ such that also condition \eqref{eq: xO condition} is satisfied. 
This completes the proof of the remaining equilibrium class and therefore of Theorem~\ref{thm:EQ characterization}. \qed

\subsection{Proof of Theorem~\ref{thm:multiplicity of EQ}}
Existence of the cost thresholds $k_B$, $k_R$ for the periodic and breakdown equilibrium, and the recovery equilibrium follows from arguments in the proof of Theorem 1 (see Equation \eqref{eq:SP rearranged} for the bound $k_B$ and the equation pair \eqref{eq: xU condition}, \eqref{eq: xO condition} for bound $k_R$.) 
 The three threshold levels in $u$ are determined as follows. For the periodic equilibrium, $u$ must be low enough so that not exerting effort at any reputation is a best response for the agent. Since the principal inspects immediately once $\bar x$ is reached, the highest incentive to exert effort is at $\bar x$. Thus, $\bar u_P$ is the largest value of $u$ such that $D_P(\bar x) \le \frac{c}{\l}$, where the subscripts $P$ indicate that $D$ is calculated with the corresponding equilibrium strategies. As inspection occurs with probability 1 at $\bar x$, we have $D_P(\bar x) = U_P(1,1)-U_P(0,0)$.
 Thus \begin{align}
 \bar u_{P} \equiv \sup \{ u >0 \ \vert \ U_P(1,1)-U_P(0,0) \le \frac{c}{\l} \}.
 \end{align}
 For $\underline u_B$ it must be the case that starting from belief $\bar x$, the agent can be incentivized to exert effort by choosing a high enough inspection rate. From the definition of the inspection rate that achieves indifference in the proof of Theorem~\ref{thm:EQ characterization} we can conclude that 
 Thus 
 \begin{align}
 \underline u_{B} \equiv \inf \{ u >0 \ \vert \ U_B(1,1)-U_B(0,0) > \frac{c}{\l} \}.
 \end{align}
 The difference between $\bar u_P $ and $\underline u_{B}$ stems from the fact that the agent's expected utility $U(p,\theta)$ is computed with different equilibrium strategy pairs.
 
 Finally, for the equilibrium with recovery, $u$ must be large enough to incentivize effort from reputation $p=0$ onward until $\underline x$ is reached and an inspection is performed. 
 Denoting, as before by $\t(0,\underline x)$ the time that it takes for the reputation to move from 0 to $\underline x$, we have that 
 $D_R(0) = e^{- r \t(0,\underline x)} D_R(\underline x)$ 
 Again, as the inspection is performed with certainty at reputation $\underline x$, the definition of $\underline u_R$ is 
 \begin{align}
 \underline u_{R} \equiv \inf \{ u >0 \ \vert \ e^{-r \t(0,\underline x)}\Paren{U_R(1,1)-U_R(0,0)} > \frac{c}{\l} \}.
 \end{align}
 In comparison to the equilibrium with breakdown, the utility difference upon inspection $U(1,1)-U(0,0)$ must now be even higher because recovery from $p=0$ requires the agent to be willing to exert effort for which he is rewarded only after the time it takes to be inspected. 
 
 \textbf{First inequality} $\underline u_B< \bar u_P$: In both equilibria $P$ and $B$, as there is no recovery we have $U_P(0,0) = 0$ and $U_B(0,0) = 0$. In the periodic inspection equilibrium, given the threshold $\bar x$, $U_P(1,1)$ follows the recursive equation
 \begin{align}
 U_P(1,1) = \frac{u}{r}\Paren{1- \bar{x}^{\frac{r}{\l}}} + \bar{x}^{\frac{r}{\l}} \bar{x} U_P(1,1),
 \end{align}
 where, since there is no effort between reputation $p=1$ and reputation $\bar x$, we have $e^{-r \tau(1, \bar x)} = \bar{x}^{\frac{r}{\l}}$.\footnote{Without effort, the belief ODE is $\dot p_t = - \l p_t$. Starting with $p_0 = 1$, the solution is $p_t = e^{- \l t}$. With $\bar x = p_{\tau(1,\bar x)} = e^{-\l \tau(1,\bar x)}$.} 
 Solving for $U_P(1,1)$ gives
 \begin{align}
 U_P(1,1) = \frac{1- \bar{x}^{\frac{r}{\l}}}{1-\bar{x}^{\frac{r+\l}{\l}}}\frac{u}{r}.
 \end{align}
 
 For the breakdown equilibrium, as the agent is indifferent between any effort choice from reputation $\bar x$ onward, we can compute $U_B(1,1)$ supposing that the agent chooses $\eta(\bar x) = \bar x$. Then, if the principal inspects at rate $\s$, the agent's expected utility satisfies 
 \begin{align}
 U_B(1,1) = \frac{u}{r}\Paren{1- \bar{x}^{\frac{r}{\l}}} + \bar{x}^{\frac{r}{\l}}\Paren{ \frac{u - \bar x c }{r+\s} + \frac{\s}{r+\s} \bar x U_B(1,1) }.
 \end{align}
 Solving for $U_B(1,1)$ gives
 \begin{align}
 U_B(1,1) = \frac{\Paren{1-\frac{\s}{r+\s} \bar{x}^{\frac{r}{\l}}}\frac{u}{r} - \frac{1}{r+\s}\bar{x}^{\frac{r+\l}{\l}} c}{1-\frac{\s}{r+\s}\bar{x}^{\frac{r+\l}{\l}} }.
 \end{align}
 Recall that we show in the proof of Theorem~\ref{thm:EQ characterization} that the threshold $\bar x $ is the same in the periodic and the breakdown equilibrium. 
The expression for $U_B(1,1)$ converges to the payoff $U_P(1,1)$ as $\s\to \infty$. 
 Further, the derivative of $U_B(1,1)$ with respect to $\s$ is 
 \begin{align}
 \frac{\del U_B(1,1)}{\del \sigma} = \frac{\bar{x}^{\frac{r}{\l}}}{ \Paren{r+\s(1-\bar{x}^{\frac{r+\l}{\l}})}^2}\Brac{-\Paren{1-\bar{x}}u +\Paren{1-\bar{x}^{\frac{r+\l}{\l}}}\bar{x} c }.
 \end{align}
 This derivative is negative whenever $u > c \frac{1-\bar{x}^{\frac{r+\l}{\l}}}{1-\bar{x}} \bar{x}$. One can show that $\frac{1-\bar{x}^{\frac{r+\l}{\l}}}{1-\bar{x}} \bar{x}$ is increasing in $\bar x$ with $\lim_{\bar x \to 1} = \frac{r+\l}{\l}$.\footnote{See end of the proof for the second inequality for details.}
 We have $u > \frac{r+\l}{\l}c$ by assumption. Thus, the utility above is decreasing in $\s$. At any inspection rate $\s \in (0,\infty)$, the agent's expected utility $U_B(1,1)$ in the random inspection equilibrium is higher than $U_P(1,1)$ in the periodic inspection equilibrium. Therefore, we have $\bar u_P >\underline u_B$ as both are defined as the value at which $U(1,1)$ in the respective equilibrium is equal to $\frac{c}{\l}$. 
 \\
 
 \textbf{Second inequality} $\underline u_B< \underline u_R$: Take any value $u > \underline u_R$ such that an equilibrium with recovery exists. We will show that a breakdown equilibrium must also exist for approval values slightly below $u$. To do this, we modify the players' strategies in the recovery equilibrium in three steps to construct a breakdown equilibrium: First, we get rid of recovery, eliminating the agent's effort and the principal's inspections after reputation $p=0$ was reached. For the second step, denote the upper inspection threshold in the recovery equilibrium by $\bar x_R$, to distinguish it from the upper threshold in the breakdown equilibrium, denoted $\bar x_B$. We change $\bar x_R$ to $\bar x_B$ and show that the latter is lower. Third, we adjust the inspection rate to the equilibrium rate in the breakdown equilibrium and show that it also lies below the original rate from the recovery equilibrium. Showing that each of these steps increases $U(1,1)-U(0,0)$ shows the desired result as it implies $U_B(1,1)-U_B(0,0) > U_R(1,1)-U_R(0,0) > e^{-( r+\l) \t(0,\underline x_R)}\Paren{ U_R(1,1)-U_R(0,0)}$. 
 
 In the recovery equilibrium with reputation cutoffs $\underline x_R$ and $\bar x_R$, the agent's expected utilities $U_R(0,0)$ and $U_R(1,1)$ satisfy
 \begin{dense}
 \begin{align}
 &U_R(0,0) = \Paren{1-(1-\underline x_R)^\frac{r}{\l}}\frac{-c}{r} + (1-\underline x_R)^\frac{r}{\l}\Paren{\underline x_R U_R(1,1)+(1-\underline x_R)U_R(0,0) }, 
 \\
 &U_R(1,1) = \frac{u}{r}\Paren{1- \bar{x}_R^{\frac{r}{\l}}} + \bar{x}_R^{\frac{r}{\l}}\Paren{ \frac{u - \bar p_R c }{r+\s} + \frac{\s}{r+\s} \Paren{\bar{x}_R U_R(1,1) + (1-\bar{x}_R)U_R(0,0)}}.
 \end{align}
 \end{dense}
 Solving the second equation for $U_R(1,1)$ gives
 \begin{dense}
 \begin{align}
 U_R(1,1) = \frac{1- \bar{x}_R^{\frac{r}{\l}}}{1-\frac{\s}{r+\s}\bar{x}_R^{\frac{r+\l}{\l}} }\frac{u}{r} + \frac{\bar{x}_R^{\frac{r}{\l}}}{1-\frac{\s}{r+\s}\bar{x}_R^{\frac{r+\l}{\l}} }\Paren{ \frac{u - \bar{x}_R c }{r+\s} + \frac{\s}{r+\s} (1-\bar{x}_R)U_R(0,0)}.
 \end{align}
 \end{dense}
 The coefficient of $U_R(0,0)$ on the right-hand side is equal to $ \frac{\s \bar{x}_R^{\frac{r}{\l}} \sigma \bar{x}_R^{\frac{r+\l}{\l}}}{r+\s-\sigma\bar{x}_R^{\frac{r+\l}{\l}} }$ and thus smaller than 1.

 First, consider a combination of strategies (not necessarily equilibrium) in which the agent never invests after $p=0$ and the principal never inspects again after $p=0$ was reached. After reaching $p=1$, both players follow the same strategies as in the recovery equilibrium. 
 The agent's continuation utility $U(1,1)$ from this strategy pair is equal to $U_R(1,1)$ above after setting $U_R(0,0) = 0$. Note that the difference $U(1,1) - U(0,0)$ after this change is larger than in the original recovery equilibrium because the coefficient of $U_R(0,0)$ is smaller than 1. 
 
 Second, adjust the inspection threshold $\bar{x}_R$ to the new principal optimum. For given values of $V(1)$ and $V(0)$, the smooth pasting condition determines $\bar{x}$ as 
 \begin{align}\label{eq:barx}
 \bar{x} = \frac{L+ r(V(0) - k)}{ H+L - (r+\l)(V(1)-V(0))}
 \end{align}
 If we treat $V(0)=w$ as a parameter, changes in $w$ have an effect on $V(1)=V(1;w)$ and on the new value $\bar{x}$. We show that the fraction on the right-hand side is increasing in $w$ 
 The recursive formulation of $V(1;w)$ is 
 \begin{align}
 V(1;w) = - \frac{L}{r}(1- \bar{x}^\frac{r}{\l}) + \frac{H+L}{r+\l}\Paren{1-\bar{x}^\frac{r+\l}{\l}} + \bar{x}^\frac{r}{\l}\Paren{(1-\bar{x}) w + \bar{x} V(1;w) - k }.
 \end{align}
 Solving for $V(1;w)$ gives
 \begin{align}
 V(1;w) = - \frac{L}{r}\frac{1- \bar{x}^\frac{r}{\l}}{1-\bar{x}^\frac{r+\l}{\l}} + \frac{H+L}{r+\l} + \frac{\bar{x}^\frac{r}{\l}}{1-\bar{x}^\frac{r+\l}{\l}}\Paren{(1-\bar{x}) w - k }.
 \end{align}
 Thus, $\frac{\del }{\del w}V(1;w) = \frac{\bar{x}^\frac{r}{\l} (1-\bar{x})}{1-\bar{x}^\frac{r+\l}{\l}}$. Consider 
 \begin{align}
 \frac{\de}{\de w} \log\Paren{ \frac{L+ r(w - k)}{ H+L - (r+\l)(V(1;w)-w)}}.
 \end{align}
 Multiplying the resulting derivative by the numerator $L+ r(w - k)$, which must be positive since $\bar x >0 $ and $H+L - (r+\l)(V(1)-V(0)) >0$ shows\footnote{See discussion following \eqref{eq:waiting optimal at bar x}.} that $\bar{x}$ is increasing in $w$ if and only if
 \begin{align}
 r - \bar{x}(r+\l)\Paren{1-\frac{\bar{x}^\frac{r}{\l} (1-\bar{x})}{1-\bar{x}^\frac{r+\l}{\l}}} >0,
 \end{align}
 or equivalently:
 \begin{align}\label{eq:xbar-inequality}
 \frac{r+\l}{r}\frac{1-\bar{x}^\frac{r}{\l} }{1-\bar{x}^\frac{r+\l}{\l}} <\frac{1}{\bar{x}}.
 \end{align}
 To show that this inequality must always hold define $\g = \frac{r+\l}{\l}$ and rewrite the inequality as 
 \begin{align}
 \Paren{\frac{\g}{\g-1}}\frac{1-\bar{x}^{\g-1}}{1- \bar{x}^{\g} }<\frac{1}{\bar x}.
 \end{align}
 For any parameter $\g>1$ we have that 
 \begin{align}
 \frac{1-p^{\g-1}}{(\g-1) \log(p)} = -\int_0^1 p^{(\g-1) t} \de t.
 \end{align}
 We then have 
 \begin{align}
 \frac{\g}{\g-1}\frac{1-\bar{x}^{\g-1}}{1-\bar{x}^{\g}} = \frac{\int_0^1 \bar{x}^{(\g-1) t} \de t}{\int_0^1 \bar{x}^{\g t} \de t}, 
 \end{align}
 Using $0<\bar{x}<1 $, we get 
 \begin{multline}
 \frac{\int_0^1 \bar{x}^{(\g-1) t} \de t}{\int_0^1 \bar{x}^{\g t} \de t} = \frac{\int_0^1 \bar{x}^{\g t} \bar{x}^{-t} \de t}{\int_0^1 \bar{x}^{\g t} \de t}\\
 = \frac{ \bar{x}^{-1} \int_0^1 \bar{x}^{\g t} \bar{x}^{(1-t)} \de t}{\int_0^1 \bar{x}^{\g t} \de t} < \frac{ \bar{x}^{-1} \int_0^1 \bar{x}^{\g t} \de t}{\int_0^1 \bar{x}^{\g t} \de t} = \frac{1}{\bar{x}}.
 \end{multline}
 This shows that as $w$ decreases to 0, the inspection threshold $\bar{x}$ must also decrease.
 After an inspection revealed high quality, the agent benefits from a lower value of $\bar{x}$. To see this, rewrite his expected utility $U(1,1)$ (with $U(0,0)=0$) as
 \begin{align}
 U(1,1) = \frac{1- \bar{x}^{\frac{r}{\l}}}{1-\frac{\s}{r+\s}\bar{x}^{\frac{r+\l}{\l}} }\frac{u}{r} - \frac{\bar{x}^{\frac{r}{\l}}}{1-\frac{\s}{r+\s}\bar{x}^{\frac{r+\l}{\l}} } \frac{u}{r}\frac{\s}{r+\s}- \frac{\bar{x}^{\frac{r+\l}{\l}}}{1-\frac{\s}{r+\s}\bar{x}^{\frac{r+\l}{\l}} } \frac{ c }{r+\s}.
 \end{align}
 The fraction $\frac{1- \bar{x}^{\frac{r}{\l}}}{1-\frac{\s}{r+\s}\bar{x}^{\frac{r+\l}{\l}} }$ decreases in $\bar{x}$ while $\frac{\bar{x}^{\frac{r}{\l}}}{1-\frac{\s}{r+\s}\bar{x}^{\frac{r+\l}{\l}} } $ and $ \frac{\bar{x}^{\frac{r+\l}{\l}}}{1-\frac{\s}{r+\s}\bar{x}^{\frac{r+\l}{\l}} }$ increase in $\bar{x}$.
 Thus, lowering the inspection threshold $\bar{x}$ increases $U(1,1)$ even further. This is intuitive: in expectation, the agent can enjoy flow utility $u$ for longer prior to being inspected. 
 
 Finally, given the new inspection threshold and the resulting value of $U(1,1)$, adjust the principal's inspection rate to achieve agent-indifference between investment levels, that is, set $\s$ to be the fix point of \eqref{eq:sigma*}. As the previous two steps both increased $U(1,1) - U(0,0)$, the new fix point lies at lower inspection rate $\s$ and higher level of $U(1,1) - U(0,0)$. Thus, since the recovery equilibrium was such that 
 \[
 e^{-(r+\l) \tau(0,\underline p_R )}(U_R(1,1) - U_R(0,0)) \ge \frac{c}{\l},
 \]
 then for the same value of $u$, the value $U_B(1,1)-0$ in the breakdown equilibrium must lie strictly above $\frac{c}{\l}$. By continuity this holds also for slightly lower $u$.
 \\
 
 \textbf{Utility thresholds are decreasing in $k$.} Define the value function of the principal as a function of $k$, i.e., $V(p,k)$ as a parameter. Denote by $\bar x^k_P$ the lower bound of the blind trust region at given inspection cost $k$. A decrease in $k$ directly decreases the principal's value, and there may be an additional increase due to a change in $\bar x^k$. Hence a decrease in $k$ has a net positive effect on $V(1,k)$. The bound $\bar x^k$ is obtained from the smooth pasting condition and given by 
 \begin{align}
 \bar x^k = \frac{L- rk}{ H+L - (r+\l)V(1,k)}
 \end{align}
 Note that the denominator is decreasing in $V(1,k)$ which itself is decreasing in $k$. Hence, the numerator is decreasing and the denominator increasing in $k$, so that $\bar x^k$ is decreasing in $k$. 
 
 Consider first the equilibrium without effort and periodic inspections, so that $\bar x^k=\bar x_P^k$. Recall that the agent's payoff in an equilibrium with periodic inspections is
 \begin{align}
 U_P(1,1) = \frac{1- \Paren{\bar x_P^k} ^{\frac{r}{\l}}}{1- \Paren{\bar x_P^k} ^{\frac{r+\l}{\l}}}\frac{u}{r}.
 \end{align}
 It is easy to see that $U_P(1,1)$ is decreasing in $ \bar x_P^k$, and thus increasing in $k$. From the definition this then implies that $\bar u_P(k)$ is decreasing in $k$. 
 
 The argument is similar for the random inspection equilibrium without recovery, where $\bar x^k=\bar p_B(k)$.As the agent is indifferent between any effort choice from reputation $\bar p_B$ onward, we can compute $U_B(1,1)$ supposing that the agent chooses $\eta(\bar p_B) = \bar p_B$. Then, if the principal inspects at rate $\s$, the agent's expected utility satisfies 
 \begin{dense}
 \begin{align}
 U_B(1,1) = \frac{u}{r}\Paren{1- \bar{p}_B(k)^{\frac{r}{\l}}} + \bar{p}_B(k)^{\frac{r}{\l}}\Paren{ \frac{u - \bar p_B(k) c }{r+\s} + \frac{\s}{r+\s} \bar p_B(k) U_B(1,1)}.
 \end{align}
 \end{dense}
 The expression is decreasing in $k$, and since $\bar p_B(k)$ is decreasing in $k$, $ U_B(1,1)$ is increasing in $k$, so that the threshold $\bar u_B$ is decreasing in $k$.  \qed
 
\subsection{Proof of Theorem~\ref{thm:self-disclosure}} Let $x_0=\max\{p|\bar \eta(p)=1\}$. On the interval $(x_0,1]$ the proof of the theorem is identical to that of cases (ii) and (iii) in Theorem~\ref{thm:EQ characterization}. It thus remains to show that neither the agent nor the principal can profitably deviate at any $p\in [0,x_0]$. At no $p\in [0,\underline x)$ is it optimal for the principal to inspect or to approve the agent since the agent exerts full effort and reports all transitions immediately. Given the agent's truthful and immediate reports at the high state, reputation jumps to zero and remains there whenever there is no report by the agent. As soon as the agent reports a change in the state for the first time, the updated reputation is given by Bayes' rule as $x_0$ at any $p\in [0,x_0)$, by construction of the agent's reporting strategy. Since $\bar\eta(x_0)=1$ by construction, it is optimal to inspect immediately at this threshold. 

Next, consider the problem of the agent. Note that a false report of a transition leads to an immediate inspection by the principal, which immediately reveals that the report was false, so that the reputation after such a report remains at zero and the agent's payoff, while in the low state, is not affected by such reports. After a transition, it is clearly optimal for the agent to report it immediately, because he is then inspected, and his reputation jumps to 1. The expected payoff from exerting full effort and reporting transitions truthfully is 
\[
U(0,0)=\frac{\lambda}{r+\lambda} U(1,1). 
\]
Moreover, by the immediate inspection after a transition, we have $U(0,1)=U(1,1)$. Therefore, the incentive differential is 
\[
D(0)= U(0,1)-U(0,0)=\frac{r}{r+\lambda} U(1,1).
\]
Note that the incentive differential is strictly increasing in $U(1,1)$ which itself is strictly increasing in $u$. Thus, for sufficiently large $u$, we have $D(0)>c$, so that maximum effort is indeed optimal. \qed

\subsection{Proof of Proposition~\ref{prop:value comparison}}
(1.) $V_R(p)\ge V_B(p)$. By the proof of Theorem~\ref{thm:multiplicity of EQ}, the inspection cutoff $\bar x$ in the recovery equilibrium is larger than the cutoff in the breakdown equilibrium. Thus, for any reputation, and also as a function of time since the last inspection, the agent exerts higher effort in the recovery than in the breakdown equilibrium. Suppose now the principal uses her inspection strategy from the breakdown equilibrium but the agent uses the effort strategy from the recovery equilibrium. The principal would clearly be better off than in the breakdown equilibrium. Since the principal's recovery-equilibrium strategy is a best response, she gets a higher payoff in the recovery equilibrium.

(2.) $V_D(p)\ge V_R(p)$. We prove this inequality in several steps. First, we write the principal's equilibrium payoff in a general form as a function of reputation thresholds $\underline x$ and $\bar x$ and a parameter $\psi$ that represents the principal's loss from discounting. We then show that this parameter is larger for an equilibrium with disclosure, and that this increases the principal's payoff. 
\begin{enumerate}[(a)]
\item Suppose the principal's recovery equilibrium payoff at $p=0$ is given by the function
\begin{align}\label{eq:w0}
 w_0(\psi,z,w_1)=\psi (z w_1-k), 
\end{align}
Using \eqref{eq:W(p)}, we can then write the principal's recovery equilibrium payoff at $p=1$ as
\begin{dense}
 \begin{align}\label{eq:V1}
W(\psi, z,\bar x)&=\frac{H+L}{r+\lambda}+\frac{\bar{x}^{r / \lambda}}{1-\bar{x}^{\frac{r+\lambda}{\lambda}}}((1-\bar{x}) w_0(\psi,z,w_1)-k){-}\left(\frac{L}{r}\right)\frac{1-\bar{x}^{r / \lambda}}{1-\bar{x}^{\frac{r+\lambda}{\lambda}}}.
\end{align}
\end{dense}
Let $W_0(\bar x)$ denote the special case when $w_0=0$. Note that if $\bar x_B$ is the lower bound of the blind-trust region in a breakdown equilibrium, then $W_0(\bar x_B)=V_B(1)$. Therefore, $W_0(\bar x_B)\ge W_0(\bar x)$ for all $\bar x\in [0,1]$. 

\item If we set $w_1=W(\psi, z,\bar x)$ in \eqref{eq:V1} and solve for $W(\psi, z,\bar x)$, we obtain
\begin{align}\label{eq:V1b}
W(\psi, z,\bar x)=W_0(\bar x)+\frac{\left(\frac{\bar{x}^\gamma}{1-\bar{x}^\gamma}\left(\frac{1-\bar{x}}{\bar{x}}\right) \psi z\right)}{\left(1-\left(\frac{\bar{x}^\gamma}{1-\bar{x}^\gamma}\right)\left(\frac{1-\bar{x}}{\bar{x}}\right) \psi z\right)} \frac{1}{z}\left(z W_0(\bar x)-k\right)
\end{align}
where $\gamma=1+r/\lambda$. Note that 
\[
\frac{\left(\frac{\bar{x}^\gamma}{1-\bar{x}^\gamma}\left(\frac{1-\bar{x}}{\bar{x}}\right) \psi z\right)}{\left(1-\left(\frac{\bar{x}^\gamma}{1-\bar{x}^\gamma}\right)\left(\frac{1-\bar{x}}{\bar{x}}\right) \psi z\right)}
\]is increasing in $\psi z$.

\item Using \eqref{eq:W(p)}, the principal's payoffs in a recovery equilibrium, after an inspection that reveals a low state, is 
\[
V_R(0) = 
\frac{(1-\underline{x}_R)^\frac{r}{\l}}{1- (1-\underline{x}_R)^\frac{r+\l}{\l}}\Paren{\underline{x}_RV_R(1) - k},
\]
where $\underline x_R$ denotes the boundary of the blind-distrust region in the recovery equilibrium. Define
\[
\psi_R=\frac{(1-\underline x_R)^\frac{r}{\l}}{1- (1-\underline x_R))^\frac{r+\l}{\l}}.
\]

\item The principal's payoff in the equilibrium with disclosure as described in Theorem~\ref{thm:self-disclosure}, after an inspection that reveals a low state, is\footnote{This comes from the recursive expression 
$$V_D(0) = \int_0^\infty e^{-r t} e^{-\left(\l + \l \frac{1-x_0^D}{x_0^D}\right)t}\left(\l + \l \frac{1-x_0^D}{x_0^D}\right) \Phi(x_0^D),
$$ 
where $\l + \l \frac{1-x_0^D}{x_0^D}$ is the rate at which the agent reports a transition which triggers an inspection at the posterior reputation $x_0^D$.} 
\[
V_D(0) = \frac{\l / x_0^D}{r+ \l}\Paren{x_0^D V_D(1)-k}.
\]
where $x_0^D$ is the principal's posterior after a reported increase in the state. Define
\begin{align}
\psi_D= \frac{\l / x_0^D}{r+ \l}.
\end{align}

\item Next, we show that $\psi_Dx_0^D>\psi_R\underline x_R$. This inequality is equivalent to
\[
\frac{\l}{r+ \l}>\frac{(1-\underline{x}_R)^\frac{r}{\l}}{1- (1-\underline{x}_R)^\frac{r+\l}{\l}}\underline{x}_R.
\]
Rewrite the last inequality as 
$$
 \frac{1- \underline x_R}{1-(1-\underline x_R)}> \frac{(1-\underline{x}_R)^\frac{r+\l}{\l}}{1- (1-\underline{x}_R)^\frac{r+\l}{\l}} \frac{r+ \l}{\l } . 
$$
This last inequality holds because $\frac{y^{\g}}{1-y^{\g}}\g$ is strictly decreasing in $\g$ for $y \in (0,1)$ and $\g>1$.

\item We show $\underline x_RW_0(\bar x_R)-k\ge 0$. Suppose otherwise. We would then have $V_R(1)=W(\psi_R, \underline x_R,\bar x_R)< W_0(\bar x_R)\le W_0(\bar x_B)=V_B(1)$, contradicting the fact that $V_R(1)\ge V_B(1)$ as shown in Part (1.) of this proof. 

\item Finally, using the previous observations, we obtain the following chain of inequalities: 
\begin{align*}
V_D(1)&=W_0(\bar x_D)+\frac{\left(\frac{\bar x_D^\gamma}{1-\bar x_D^\gamma}\left(\frac{1-\bar x_D}{\bar x_D}\right)\psi_D x_0^D\right)}{\left(1-\left(\frac{\bar x_D^\gamma}{1-\bar x_D^\gamma}\right)\left(\frac{1-\bar x_D}{\bar x_D}\right) \psi_D x_0^D\right)} \left( W_0(\bar x_D)-\frac{k}{x_0^D}\right)\\
&\ge W_0(\bar x_R)+\frac{\left(\frac{\bar x_R^\gamma}{1-\bar x_R^\gamma}\left(\frac{1-\bar x_R}{\bar x_R}\right)\psi_D x_0^D\right)}{\left(1-\left(\frac{\bar x_R^\gamma}{1-\bar x_R^\gamma}\right)\left(\frac{1-\bar x_R}{\bar x_R}\right) \psi_D x_0^D\right)} \left( W_0(\bar x_R)-\frac{k}{x_0^D}\right)\\
&\ge W_0(\bar x_R)+\frac{\left(\frac{\bar x_R^\gamma}{1-\bar x_R^\gamma}\left(\frac{1-\bar x_R}{\bar x_R}\right)\psi_D x_0^D\right)}{\left(1-\left(\frac{\bar x_R^\gamma}{1-\bar x_R^\gamma}\right)\left(\frac{1-\bar x_R}{\bar x_R}\right) \psi_D x_0^D\right)} \left( W_0(\bar x_R)-\frac{k}{\underline x_R}\right)\\
&\ge W_0(\bar x_R)+\frac{\left(\frac{\bar x_R^\gamma}{1-\bar x_R^\gamma}\left(\frac{1-\bar x_R}{\bar x_R}\right)\psi_R \underline x_R\right)}{\left(1-\left(\frac{\bar x_R^\gamma}{1-\bar x_R^\gamma}\right)\left(\frac{1-\bar x_R}{\bar x_R}\right) \psi_R \underline x_R\right)} \left( W_0(\bar x_R)-\frac{k}{\underline x_R}\right)\\
&= V_R(1)
\end{align*}
 The first inequality holds because $\bar x_D$ is the optimal stopping threshold in the disclosure equilibrium. The second inequality follows because $x_0^D\ge p^\dagger\ge \underline x_R$. The third inequality follows because $\psi_Dx_0^D\ge \psi_R\underline x_R$. Thus, it follows that $V_D(1)\ge V_R(1)$ as required. 
 \item Using the arguments above, we have that $V_D(1)\ge V_R(1)$ implies that $V_D(0)\ge V_R(0)$, and then applying the same argument as in Part (1.) of this proof, it follows that $V_D(p)\ge V_R(p)$ for all $p\in [0,1]$.  \qed
\end{enumerate}

\newpage
\section{Online Appendix}

\subsection{Admissible beliefs}\label{subsec:BeliefRestriction}
As pointed out by \cite{klein2011negatively}, there exist Markov strategies $\eta$ for the agent for which the belief process defined by the ODE
\begin{align*}
 \dot p_t = \lambda (\tilde \eta(p_t)-p_t)
\end{align*}
admits no solution. Following \cite{board2013reputation}, we impose the following restriction on the believed strategy $\tilde \eta$:
Believed strategy $\tilde \h$ is admissible if there is a finite number of cutoffs $0\le p^1 < \cdots < p^n \le 1$, such that (i) $\tilde \h$ is Lipschitz-continuous on any interval $[0,p^1)$; $(p^i,p^{i+1})$ for $i \in \{1,\dots, n-1\}$; and $(p^{\s},1]$, and (ii) at every cutoff, $\tilde \eta$ satisfies one of the following three conditions
\begin{enumerate}
 \item $\tilde \eta(p^i)-p^i = 0$
 \item $\tilde \eta(p^i)-p^i > 0$ and $\tilde \eta$ is right-continuous at $p^i$
 \item $\tilde \eta(p^i)-p^i < 0$ and $\tilde \eta$ is left-continuous at $p^i$.
\end{enumerate} 
These restrictions are placed on the agent's believed effort strategy so that admissibility does not restrict potential deviation by the agent. These restrictions insure that the reputation ODE has a solution. 
In our case, the only possible starting points of each belief trajectory are $p_0=1$ and $p_0 = 0$, depending on the outcome of the last inspection. Since $\tilde \eta \in [0,1]$, any trajectory must remain in $p_t [0,1]$ so that the unique local solutions implies by admissibility also apply global uniqueness from both initial points.

\subsection{Recursive formulation}\label{Apx:fixed point}
\begin{lemma}
  Take the operator

  \begin{align*}
      \GG  V  (p) \equiv \sup_{F \in \D(\real_+)} \E_{\tau \sim F} \Brac{ \int_0^\t e^{-r t} v(p_t)^+ \de t\ + \ e^{-r \t} \Phi(p_\tau ; V )}, \quad \text{ where } \dot p_t \text{ as in } \eqref{eq:ReputationODE}, \text{ with } p_0 = p.
  \end{align*}
  The principal's value function is the unique bounded function $V$ satisfying $V = \GG V$; i.e. the unique fixed point of $\GG$.
\end{lemma}
Recall that the inspection function definition $$\Phi(p; V) = p V(1)+(1-p)V(0) - k. $$
In comparison to the main text, we make the dependence of $\Phi$ on the post-inspection continuation values $V$ explicit.  
\begin{proof}
The proof follows from arguments in \citet[Chapter 54]{davis1993}.
By our admissibility conditions on $\tilde \eta$, every starting point $p_0=p$ leads to a unique continuous trajectory $\{p_t\}_{t\ge 0}$ determined by the ODE \eqref{eq:ReputationODE}. 

Suppose the principal cannot inspect at all. Then her payoff is 
$$
V^{0}(p) = \int_0^\infty e^{-r t} v(p_t)^+ \de t, \quad \text{ where } \dot p_t \text{ as in } \eqref{eq:ReputationODE}, \text{ with } p_0 = p.
$$
For general $k\in \mathbb{N}$, we define $V^{k+1} = \GG V^k$, that is $V^k$ gives the value when the principal can inspect at most $k$ times. 

It is easy to see that for all $p\in [0,1]$ and all $k$ we have $0 \le V^k(p) \le H/r$, where $H/r$ would be the discounted value from continuously approving with a guaranteed high state. 
Further, $\Phi$ is a linear operator that maps any bounded function $V$ into a continuous bounded function. 

Proposition 54.18 in \citet[p. 235]{davis1993} establishes that 
$$
\lim_{k\to \infty} V^k (p) = V(p) = \sup_{\{\tau_i\}_{i=1}^\infty}\Big\{ \E \Brac{\int_0^\infty e^{-r t} v(p_t)^+ \de t - \sum_{i=1}^\infty e^{-r \t_i}k  } \Big\},
$$
where the belief trajectory starts at $p_0 = p$, evolves according to \eqref{eq:ReputationODE} between inspections and jumps to state $\theta_{\t_i}$ at the time of the $i$'th inspection. 

Finally, Theorem 54.19 in \citet[p. 236]{davis1993} confirms that the value function $V$ is the unique bounded function that satisfies the fixed-point property $V = \GG V$. 
\end{proof}
\begin{remark}
Note that the result statements in \citet[Chapter 54]{davis1993} make reference to deterministic intervention times $\{\tau_i\}$ only. 
However, holding fixed the believed agent-behavior $\tilde \eta$, there is no gain from randomization for the principal \citep[see][Chapter 52 on randomized stopping]{davis1993}. In fact, the proof establishing sufficiency of our HJB equation (which will be equivalent to the ``Quasi-Variational Inequalities'' in the terminology of \citet{davis1993}) explicitly consider randomized interventions to prove optimality through the so-called penalty method.
\end{remark}
\begin{remark}
    It is worthwhile to note that we cannot conclude that $V$ is continuous in $p$ on all of $[0,1]$ when considering arbitrary admissible $\tilde \eta$. As $\tilde \eta$ may feature discontinuities, trajectories starting from neighboring values of $p$ may take very different routes. The admissibility condition on $\tilde \eta$ guarantee, however, that paths $t \mapsto V(p_t)$ between inspections are continuous.
\end{remark}

Finally Theorem 54.28 in \cite[p. 242]{davis1993} establishes that that the value function $V$ is the unique function with Lipschitz-continuous time paths $t \mapsto V(p_t)$ satisfying 
\begin{align*}
   0=  \max \Big\{ \UU V(p) - r V(p) + v(p)^+ \ ; \ \Phi(p; V) - V(p) \Big\} , 
\end{align*}
where $\UU$ denotes the generator\footnote{See \citeauthor{davis1993}, \citeyear{davis1993}, pp. 27-33.}
 which for our case of deterministic trajectories between inspections simplifies to the time derivative $\frac{\de}{\de t} V(p_t)$, that is, $\lim_{\e \downarrow 0 } \frac{V(p_{t+\e}) - V(p_t)}{\e}$, where $\phi(s,p)=p_s$ denotes the solution to \eqref{eq:ReputationODE} with initial value $p$. 
 If $V$ is differentiable, this is obviously $V'(p)\l(\tilde \eta(p) - p)$, which gives precisely the HJB equation \eqref{eq: Principal HJB general}
in the main text. 
This condition implies the smooth-pasting condition $V'(p) = \Phi'(p)$ at any $p$ at which we transition from waiting being optimal to inspection becoming optimal.
To see this, note that $\Phi$ is a linear function in $p$ for any values of $V(0)$ and $V(1)$. In particular, $\Phi$ must be continuously differentiable. Now suppose smooth pasting is violated at a point $p_t$ satisfying $V(p_{t-\epsilon}) > \Phi(p_{t-\epsilon})$ for $\e>0$ small enough and $ V(p_{t}) = \Phi(p_{t})$. 
Then 
$$
\lim_{\e\downarrow 0}\Paren{ \frac{V(p_{t}) - V(p_{t-\e})}{\e}  - \frac{\Phi(p_{t}) - \Phi(p_{t-\e})}{\e}} \ne 0,
$$
for some $p_t$ with  $0=\UU V(p_t) - r V(p_t) + v(p_t)^+ \ = \ \Phi(p_t; V) - V(p_t)  $.
Since we transition from waiting being strictly optimal at $t-\e$ into inspection being optimal at $t$, the limit must be negative. 
This implies the inequality
\begin{align*}
    0 &= \lim_{\e\downarrow 0} \frac{V(p_{t}) - V(p_{t-\epsilon})}{\e} - r V(p_t) + v(p_t)^+ 
    \\
    & < \lim_{\e\downarrow 0} \frac{\Phi(p_{t}) - \Phi(p_{t-\epsilon})}{\e} - r \Phi(p_t) + v(p_t)^+ 
    \\
    & =   \lim_{\e\downarrow 0} \frac{\Phi(p_{t+\epsilon}) - \Phi(p_{t})}{\e} - r \Phi(p_t) + v(p_t)^+ = \frac{\de}{\de t} \Phi(p_t) - r\Phi(p_t) +v(p_t)^+, 
\end{align*}
where the equality at the beginning of the last line holds because $\Phi$ is continuously differentiable and the admissibility restrictions on $\tilde \eta$ imply that for all $t>0$, the reputation paths satisfy $\lim_{\e\downarrow 0}\frac{p_{t}-p_{t-\e}}{\e} =\lim_{\e\downarrow 0}\frac{p_{t+\e}-p_{t}}{\e}$ if we are not inspecting at $t$. 
However, the inequality $\frac{\de}{\de t} \Phi(p_t) - r\Phi(p_t) +v(p_t)^+ > 0 $ violates the quasi-variational inequality; waiting for an additional instant would strictly better that inspecting at $t$. 

\subsection{Auxiliary Results and Proofs}\label{Apx:Aux}
The first lemma below states that after every positive inspection outcome there the time until the next inspection is bounded away from zero. This is intuitive as the principal learns virtually no information from an inspection that is performed at a reputation arbitrarily close to the certainty beliefs $0$ or $1$. Thus, investing the fixed inspection cost $k$ cannot be optimal. 
\begin{lemma}\label{lm:blind regions}
In any equilibrium, $\bar x <1$ and $\underline x > 0$. 
\end{lemma}
\begin{proof} Suppose that, contrary to the first claim, $\bar x = 1$ is the supremum of all reputations at which (possibly random) inspections take place. First, it cannot be the case in equilibrium that the agent exerts maximum effort at $p=1$. If that was the case, then the principal's equilibrium posterior would remain at 1 forever, and it would not be optimal to ever inspect, in which case the agent would prefer to shirk. Therefore, the agent must exert less than maximum effort in an open neighborhood of $p=1$ in equilibrium, and the agent's reputation must fall below 1. If it is an optimal strategy for the principal to inspect at a reputation arbitrarily close to 1, then with positive probability an inspection must take place after arbitrarily small delay $\Delta$. By optimality of the principal's inspection strategy, the value function for the principal at reputation $p_t=1$ then must satisfy
\begin{align}\label{eq:blind trust integral}
V(1)=\int_t^{t+\Delta} e^{-r(s-t)}v(p_{s})ds+e^{-r \Delta}(p_{t+\Delta} V(1)+(1-p_{t+\Delta})V(0)-k).
\end{align}
Note that in the absence of an inspection, reputation evolves continuously, and therefore, we have $p_{t+\Delta}\to 1$ for $\Delta\to 0$. But then the right-hand side above is equal to $V(1)-k+o(\Delta)$, so that \eqref{eq:blind trust integral} implies $V(1)=V(1)-k+o(\Delta)$ which yields a contradiction for $\Delta\to 0$. Therefore, there is a threshold $\bar x$ bounded away from 1 such that no inspections take place on the interval $(\bar x ,1]$.

Similarly, it must be that $\underline x >0$: 
suppose $\underline x=0$. If in equilibrium $\eta(0)=0$, then it is optimal for the principal to never inspect. Suppose the agent exerts positive effort in an open neighborhood of $p=0$ in equilibrium, and the agent's reputation must rise above 0. If it is an optimal strategy for the principal to inspect at a reputation arbitrarily close to 0, then inspections must take place with arbitrarily small time delay $\Delta$. By optimality of the principal's inspection strategy, the value function for the principal at reputation $p_t=0$ must then satisfy
\begin{align}
V(0)=e^{-r \Delta}(p_{t+\Delta} V(1)+(1-p_{t+\Delta})V(0)-k).
\end{align} 
For $\Delta\to 0$, the right-hand side above is equal to $V(0)-k+o(\Delta)$, so that the equation implies $V(0)=V(0)-k+o(\Delta)$ which yields a contradiction for $\Delta\to 0$. 
\end{proof}

The next lemma shows that the agent exerts no effort at any reputation above the threshold $\bar x$.
\begin{lemma}\label{lm:no effort at top}
In any equilibrium, $\eta(p) = 0 $ for all $p\in (\bar x,1]$
\end{lemma}
\begin{proof}
Note that we must have $\eta(p) - p <0 $ for all $p> \bar x$. If this were not the case, so that $\eta(p) - p \ge 0$ for some reputation in $( \bar x , 1]$. Then there exists reputation $p' \in ( \bar x , 1]$ such that after an inspection revealed the high state, the reputation stays above $p'$ forever. Since there are no inspections above $\bar p$, the agent would then optimally never exert any effort; a contradiction to $\eta(p) - p \ge 0$. Therefore, at any $p\in ( \bar x,1]$ we have $\eta(p)<p$ and $\lim_{p\downarrow \bar x}\eta(p)\le \bar x$. Denote by $\tau(p,p')$ the finite time required for the agent's reputation to fall from $p$ to $p'<p$, given effort strategy $\eta$. 

As there are no inspections on $( \bar x,1]$, a difference in the agent's payoffs across states arises only once reputation reaches $\bar x$, so that
\[
D(p)=\lim_{p'\downarrow \bar x}e^{-(r+\l)\tau(p,p')}D(p'). 
\]
First, if we had $\lim_{p'\downarrow \bar x}D(p') \le 0$, then also $D(p) \le 0$ for all $p>\bar x$, and the agent exerts no effort at any such belief. 
Second, if we have $\lim_{p'\downarrow\bar x}D(p') > 0$, then 
\[
\lim_{p'\downarrow \bar x}e^{-(r+\l)\tau(p,p')}D(p') < \lim_{p'\downarrow \bar x}D(p').
\]
Moreover, since $\lim_{p'\downarrow \bar x}\eta(p')\le \bar x<1$, the agent cannot have a strict incentive to exert effort in the limit, i.e.,
\[
\lim_{p'\downarrow \bar x}D(p')\le\frac{c}{\lambda}.
\]
This implies that $D(p) <\frac{c}{\l}$ for all $p>\bar x_1$ which proves that $\eta(p) = 0$ for all $ p \in (\bar x , 1]$. 
\end{proof}

Next, suppose that the inspection at reputation $\bar x$ is performed with probability 1. This can occur in equilibrium only if the agent exerts zero effort almost everywhere as the following result shows. 
\begin{lemma}\label{lm: immediate inspection at top implies no effort}
Consider an equilibrium in which the principal inspects immediately at reputation $\bar x$. Then $U(1,1)-U(0,0)\le \frac{c}{\l}$, where $U(p, \th)$ is the agent's continuation utility at reputation $p$ in state $\th$.
\end{lemma}
\begin{proof}
By Lemma~\ref{lm:no effort at top} we have $\eta(p) = 0 $ for all $p> \bar x$. If the inspection at $\bar x$ occurs with probability one, this requires that $U(1,1)-U(0,0)\le \frac{c}{\l}$: If the current reputation is $p>\bar x$, then the agent's incentive to exert effort, given by $D(p)$, is equal to $e^{-(r+\l)\tau(p, \bar x)} \Paren{U(1,1)-U(0,0)},$
where $\tau(p, \bar x)$ is the time the reputation takes to decline from $p$ to $\bar x$. 
If we had $U(1,1)-U(0,0)> \frac{c}{\l}$, then at some $p$ above but arbitrarily close to $\bar x$ we would have $D(p) > \frac{c}{\l}$ as $\tau(p,\bar x)$ is arbitrarily close to 0 as $p \to \bar x$. Then, $D(p)$ would be strictly above $\frac{c}{\l}$, which would prescribe full effort from the agent; contradicting $\eta(p) = 0 $ for all $p> \bar x$. 
\end{proof}

\begin{lemma}\label{lm:random inspection at top implies rate}
Consider an equilibrium in which the principal does not inspect immediately at reputation $\bar x>0$. Then either the equilibrium is outcome equivalent to the equilibrium with immediate inspection or the inspection at $\bar x$ arrives at a rate over time $\s(p) > 0$ for an open interval $(\bar x - \e, \bar x)$.
\end{lemma}
\begin{proof}
Recall from Lemma~\ref{lm:no effort at top} that the agent exerts no effort on $(\bar x, 1]$. 
This implies that $\lim_{p\searrow \bar x} D(p) \le c/\l$. 
If there was an atom in the inspection probability distribution at $\bar x$, then $D(p)$ would jump downward to $D(\bar x)$ strictly below $c/\l$ once the atom is passed, and the agent would strictly prefer to shirk.
The arguments in the previous case without effort show that ---given that she is willing to inspect at $\bar x$--- the principal strictly prefers to inspect at reputations arbitrarily close to the left of $\bar x$ when the agent exerts no effort. This would lead to an equivalent action and reputation path to the one where an immediate inspection is performed at reputation $\bar x$ itself.
If there is no atom at $\bar x$ and $\bar x$ is the supremum of reputations at which $\s(p) >0$, the principal must inspect with positive hazard rate on some neighbourhood $(\bar x - \epsilon, \bar x)$.
\end{proof}

The principal's indifference condition between inspecting and not inspecting fixes the agent's effort.
When this effort level is interior, the agent's indifference condition in turn fixes the inspection rate of the principal. The next result shows this formally.

\begin{lemma}\label{lm:no gap in inspection region}
Suppose that the agent exerts no effort and it is optimal for the principal to inspect at $p=p^\dagger$ as well as $p=\bar x$. The the principal must inspect immediately at any $p \in (p^\dagger, \bar x)$.
\end{lemma}
\begin{proof}
    Suppose to the contrary that there is a non-empty interval $(p_1,p_2) \subset (p^\dagger, \bar x)$ such that it is optimal for the principal to wait inside the interval and optimal to inspect at $p_2$. Then, the first term in the max operator on the right-hand side of \eqref{eq:Principal HJB in proof} must be constantly equal to 0 on the interval $(p_1,p_2)$. The corresponding  ODE with some boundary value $V_{no}(p_1)$ has the solution 
$$
V_{no}(p) = - \frac{L}{r} + p \frac{H+L}{r+\l} + \Paren{\frac{p_1}{p}}^{\frac{r}{\l}} \Paren{ V_{no}(p_1) + \frac{L}{r} - p_1 \frac{H+L}{r+\l} }, \qquad \text{ for } p \in (p_1,p_2).
$$ 
Depending on the sign of $ V_{no}(p_1) + \frac{L}{r} - p_1 \frac{H+L}{r+\l} >0$; this function is strictly convex, (if $>0$), strictly concave (if $<0$), or linear (if $=0$) on the entire interval. 
Given that the agent exerts no effort, $\dot p_t < 0$ for all $p_t >0$. Hence, the belief transitions smoothly into  $(p_1,p_2)$ from above; implying that the value function must satisfy smooth pasting at the regular boundary $p_2$.
We can rule out that $V_{no}$ is strictly convex or strictly concave: 
If $V_{no}$ is strictly convex, the linearity of $\Phi$ and $V_{no}(p_2)=\Phi(p_2)$ imply $V_{no}(p_1) > \Phi(p_1)$, contradicting the optimality of inspecting at $p_2$. 
If $V_{no}$ is strictly concave, we get $V_{no}(p) < \Phi(p)$ for all $p\in(p_1, p_2)$, making inspection strictly optimal on the interval. 

In the remaining case that $ \Paren{ V_{no}(p_1) + \frac{L}{r} - p_1 \frac{H+L}{r+\l} }=0$, would imply that 
$V_{no}(p) = p \frac{H+L}{r+\l} - \frac{L}{r}$, or, equivalently,
$$
V_{no}(p) = p \Paren{\frac{H}{r+\l} + \frac{\l}{r+\l} \frac{-L}{r}} +(1-p)\frac{-L}{r}.
$$

This cannot be optimal because the right-hand side of the above expression denotes the expected discounted payoff, in state $p$, from approving forever onward given that the agent exerts no effort. The principal does strictly better than that by refusing approval whenever $p<p^\dagger$.
\end{proof}

\begin{lemma}\label{lm:random inspection at top implies mixed effort and constant inspection rate}
Consider an equilibrium in which the principal inspects randomly at reputation $\bar x$. Let $x^* = \sup\{ p \in [0,1] \vert \tilde \eta(p) \ge p \}$ be the lowest reputation which can be reached from above while no inspection is conducted when the believed effort strategy is $ \tilde \eta$. Then the agent's effort on $[x^*, \bar x]$ is given by the linear function \begin{align*}
 \eta(p) = \frac{x^*(\bar x - p)}{\bar x - x^*}, 
\end{align*}
and the principal inspects at constant hazard rate $\s^* > 0$ 
\end{lemma}
\begin{proof} We first focus on the agent's effort and then turn to the principal's inspection rate. 
Consider the principal's HJB from \eqref{eq: Principal HJB general} when the believed effort strategy is $\tilde \eta$:
\begin{align} \label{eq:HJB proof part B}
 0 = \max \left\{ v(p)^+ + \l V'(p)\Paren{\tilde \eta(p) -p} - r V(p)   ;   \Phi(p)- V(p) \right\}.
\end{align}
The principal is willing to mix between inspecting and waiting if and only both terms in the max operator are 0. 
Whenever this is the case, then 

the first term is strictly increasing in $\tilde \eta(p)$. The principal is indifferent between inspecting and waiting when $ \tilde \eta (p) = \bar \eta(p)$ given in \eqref{eq:eta bar} and which we restate here: 
\begin{align} \label{eq:eta bar proof}
 \bar \eta(p) = \frac{r\Paren{V(0)-k}}{\l \Paren{V(1)-V(0)}} + p \frac{r+\l}{\l} - v(p)^+ \frac{1}{\l \Paren{V(1)-V(0)}}. 
\end{align}
The optimality of inspecting at $\bar x$ for the first time implies that 
\begin{align*}
 \bar x= \sup\{ p \in [0,1] \ \lvert \ \bar \eta(p) = 0 \}.
\end{align*}
Recall that $v(p) = -L + p(H+L)$. The first term in the max operator in \eqref{eq:HJB proof part B} at $\bar x$ with $\tilde \eta(\bar x) = 0$ is 
\begin{dense}
\begin{align}\label{eq:waiting optimal at bar x}
 \Paren{-L + \bar x (H+L)}^+ - \l \bar x \Paren{V(1)-V(0)} - r \Paren{ V(0)+ \bar x \Paren{V(1)-V(0)}-k},
\end{align}
\end{dense}
where we make use of the second term in the HJB being 0; i.e. $V(\bar x)= \Phi(\bar x)= V(0)+ \bar x \Paren{V(1)-V(0)}-k$.
The term in \eqref{eq:waiting optimal at bar x} being equal to 0 implies that $\bar x > p^\dagger$ as well as $(r+\l)\Paren{V(1)-V(0)} < H+L$.
It follows that $ \bar \eta(p)$ is linearly increasing for $p<p^\dagger$ and decreasing for $p > p^\dagger$. The maximum is attained at $p^\dagger = \frac{L}{H+L}$.

Since the principal mixes on the interval $(\bar x - \varepsilon, \bar x)$, the agent must be exerting effort $\bar \eta (p)$, which lies strictly in $(0,1)$ for small $\varepsilon$. By \eqref{eq:best-response}, interior effort is optimal for the agent only if $D(p)= \frac{c}{\l}$ on this interval. Since the inspection is not immediate, this requires that $U(1,1)-U(0,0) > \frac{c}{\l}$. 
This implies that if the principal were to inspect immediately at some reputation $p$, then the agent would exert full effort at reputations that are passed just before reaching $p$. As a result, an immediate inspection cannot occur at any reputation that is reached from above in equilibrium. Similarly, when $D(p)= \frac{c}{\l}$ on an interval $(\bar x - \varepsilon, \bar x)$, if the principal would cede to inspect at some reputation, $D(p)$ would fall below $\frac{c}{\l}$, so that the agent would optimally exert no effort and, since $\bar \eta (p) > 0$, the principal would then strictly prefer to inspect immediately; a contradiction. 
Thus the principal must inspect randomly starting at reputation $\bar x$ at least as long as the reputation transitions downward. Since randomization is optimal for the principal if and only if the agent chooses effort $\bar \eta (p)$, the reputation moves downward as long as $\bar \eta(p)-p <0$. 
Let $x^* = \sup\{ p \in [0,1] \lvert \ \bar \eta(p) = p \}$ be the reputation at which $\bar \eta(p)$ crosses $p$ from above (see Figure~\ref{fig:eta}). While the agent exerts effort $\bar \eta$ for reputations in $(x^*,\bar x]$, as long as no inspection has arrived, the reputation declines gradually and approaches $x^*$ from above. By the arguments above, the reputation cannot gradually decline strictly below $x^*$ because this would require effort lower than $\bar \eta$ which would make an immediate inspection strictly optimal for the principal. 

Now we establish the equilibrium inspection rate that keeps the agent indifferent between any effort level for reputations in $(p^*, \bar x)$. Setting $\frac{\de}{\de t} D(p_t) = 0 $ and $ D(p_t) = \frac{c}{\l}$ in \eqref{eq:IncentiveDiff} gives
\begin{align*}
0 = (r+\l) \frac{c}{\l} + \s(p_t) \frac{c}{\l} - \s(p_t)\Brac{U(1,1)-U(0,0)}.
\end{align*}
Solving for $\s(p_t)$ shows that 
$D(p_t)$ is constant and equal to $\frac{c}{\l}$ if and only if the principal inspects at rate $\s^*$ solving 
\begin{align} \label{eq:inspection rate fixed point}
\s=\frac{ (r+\lambda ) c/\lambda}{ U_\s(1,1)-U_\s(0,0)-c/\lambda}.
\end{align}
We write $U_\s(1,1)$ and $U_\s(0,0)$ to highlight that the agent's equilibrium continuation utilities after inspection are dependent on the inspection rate so that condition \eqref{eq:inspection rate fixed point} is a fix point problem.
The confirmation that a solution $\s^*$ to the fix point condition \eqref{eq:inspection rate fixed point} exists follows in Lemma~\ref{lm:fix point inspection rate}.
\end{proof}

\begin{lemma}\label{lm:fix point inspection rate}
Take any equilibrium in which the agent exerts positive effort at some reputation. 
There exists a value $\s^*>0$ such that the agent is indifferent between any effort level in $[0,1]$ if the next inspection arrives at constant rate $\s^*$.
\end{lemma}
\begin{proof} We first take as given a constant rate of inspection $\s$ and suppose that the agent is indifferent between any effort level with this rate to compute the expected payoffs of the agent, 
In the second step, we use the resulting values for $U_\s(0,0)$ and $U_\s(1,1)$ to show that Equation \eqref{eq:inspection rate fixed point} indeed has a fix point $\s^*$. 

If the initial reputation is $p_0 = p \in (x^*,\bar x)$ and the current state is $\th_0 = 0$, the agent's expected payoff is
\begin{multline*}
U_\s\left( p,0\right)=\int_0^\infty e^{-\left(r+\lambda +\s\right) t} \big( u- \eta(p_t)c +\s U_\s(0,0) \\+\lambda ( \eta(p_t)U_\s(p_t,1) +(1- \eta(p_t))U_\s(p_t,0)\big) \, \de t,
\end{multline*}
where $\{p_t\}_{t\ge 0}$ is the time-path of reputation following $\dot p_t=\lambda (\tilde \eta(p)-p)$ with $p_0= p$ and $\lim_{t\to \infty}p_t=x^*$. Because of the agent's indifference during the inspection phase, we have $c=\lambda(U_\s(p_t,1)-U_\s(p_t,0))$ for all $t$, and therefore, the agent's expected payoff is
\[
U_\s\left(p_0,0\right)=\int_0^\infty e^{-\left(r+\lambda +\s\right) t} \big(ru+\s U_\s(0,0) +\lambda U_\s(p_t,0)\big) \, \de t
\]
Note that the integral depends on $p_t$ only through $U_\s(p_t,0)$, so that $U_\s(p_t,0)=U_\s(p_0,0)$. We can thus solve explicitly:
\[ 
U_\s(p_0,0)=\frac{1}{r+\s}u+\frac{\s}{r+\s}U_\s(0,0)
\]
Similarly, the expected payoff for the agent at high quality is 
\[
U_\s\left( p_0,1\right)=\int_0^\infty e^{-\left(r+\lambda +\s\right) t} \left(u+\s U_\s(1,1) +\lambda U_\s( p_0,0)\right) \, \de t
\]
which when solving the integral becomes
\[ U_\s(p_0,1)= \frac{1 }{r+\s+\lambda }u+\frac{\lambda}{r+\s+\lambda }U(p_0,0)+\frac{\s }{r+\s+\lambda } U(1,1) 
\]
It is straightforward to check that indeed
\[
U_{\bar \s}(p_0,1)-U_{\bar \s}(p_0,0)=\frac{c}{\lambda},
\]
so from Equation \eqref{eq:best-response}, effort $ \eta(p_0)$ is a best response for the agent. 

Now, the agent's payoff at reputation $p=1$ is 
\begin{multline*}
U_\s(1,1)=\int_0^{T_1} e^{s (-(r+\l ))} \left( u+\lambda \left( \left(1-e^{-r (T_1-s)}\right)u+ e^{-r (T_1-s)}U_\s(\bar p,0)\right)\right) \, ds\\+U(\bar p,1) e^{T_1 (-(r+\l ))}.
\end{multline*}
Since $U_\s(p,\theta)\in (-c,u)$ the expected payoff for the agent $U_\s(1,1)$ lies within the range $(U_\s(0,0),u)$. We can then solve for $U_\s(1,1)$:
\begin{multline}\label{eq:U11 solved}
U_\s(1,1)=\frac{1}{\s+r} \left(1-e^{-T (r+\lambda )}\right) u\\
+\frac{\s}{\s+r} \left(1-e^{-T r}\right) u+e^{-T (r+\lambda )} \frac{1}{\s+r+\lambda } u
+e^{-T r} \frac{\s}{\s+r} U_\s(0,0)\\-e^{-T (r+\lambda )} \frac{\s}{\s+r} U_\s(0,0) +e^{-T (r+\lambda )} \frac{\s}{\s+r+\lambda } U_\s(1,1)\\
+e^{-T (r+\lambda )} \frac{\lambda }{\s+r+\lambda
 } \left(\frac{1}{\s+r} u+\frac{\s}{\s+r} U_\s(0,0)\right)
 \end{multline}
The continuation payoff for the agent starting at the initial belief of the inspection phase $p=0$. The expected payoff for the agent at low quality is
\begin{multline}
U_\s(0,0)=\int_0^{T_0} e^{-(r+\lambda ) s} \left(- c+\lambda \left(\left(1-e^{-r (T_0-s)}\right)
 (-c)+e^{-r (T_0-s)} U_\s(1,1)\right)\right) \de s\\
 +e^{-(r+\lambda ) T_0} U_\s(0,0)
\end{multline}
Since $U_\s(p,\theta)\in (-c,u)$ the expected payoff for the principal $U_\s(0,0)$ lies within the range $(-c,U_\s(1,1))$.
\begin{dense}
\begin{align}\label{eq:U00 solved}
U_\s(0,0)=\left(1-e^{-T_0 r}\right) (-c)+e^{-T_0 r} \left(\left(1-e^{-T_0 \lambda }\right)
 U_\s(1,1)+e^{-T_0 \lambda } U_\s(0,0)\right)
\end{align}
\end{dense}
Now we can solve \eqref{eq:U00 solved} for $U_\s(0,0)$
\begin{align}\label{eq:U00 as fct of U11}
U_\s(0,0)=\frac{\left(1-e^{-T_0 r}\right)}{1-e^{-T_0 (r+\l )}} (-c)+\frac{e^{-T_0 r}\left(1-e^{-T_0 \lambda }\right)}{1-e^{-T_0 (r+\l )}} 
 U_\s(1,1).
\end{align}
Note here that
\[
U_\s(1,1)-U_\s(0,0)=\left(1-\frac{e^{-T_0 r}\left(1-e^{-T_0 \lambda }\right)}{1-e^{-T_0 (r+\l )}}\right)U_\s(1,1)+\frac{\left(1-e^{-T_0 r}\right)}{1-e^{-T_0 (r+\l )}} c. 
\]
Since $U_\s(1,1)$ is strictly increasing in $u$, so is $U_\s(1,1)-U_\s(0,0)$, so that regardsless of $\s$, we have $U_\s(1,1)-U_\s(0,0)>\frac{c}{\lambda}$ for $u$ large enough. Now, substitute \eqref{eq:U00 as fct of U11} in \eqref{eq:U11 solved} to solve for $U_\s(1,1)$ as a function of $\sigma$. This shows that for any value of $\s$, there is a unique solution for $U_\s(0,0)$ and $U_\s(1,1)$ with $-c\leq U_\s(0,0)\leq U_\s(1,1)<u$. Now we need to show that there is a value $\s^*$ which solves the indifference condition 
\[
\s:=\frac{ (r+\lambda ) c/\lambda}{ U_\s(1,1)-U_\s(0,0)-c/\lambda}.
\]
Define the auxiliary function
\[
h(\s)=\frac{ U_\s(1,1)-U_\s(0,0)-c/\lambda}
{(r+\lambda)c/\lambda}-\frac{1}{\s}.
\]
Then a solution $\s^*$ to \eqref{eq:sigma*} satisfies $h(\s^*)=0$. 

Note first from \eqref{eq:U11 solved} and \eqref{eq:U00 solved} that both $U_\s(1,1)$ and $U_\s(0,0)$ are continuous in $\s$ for $\s> 0$. Therefore, $h(\s)$ is continuous. Moreover, as argued above, for given $\s>0$, and $u$ large enough, we have $U_\s(1,1)-U_\s(0,0)>\frac{c}{\lambda}$. Fix such an $u$, and let $\tilde \sigma$ be the infimum of all values of $\s$ for which 
$$U_\s(1,1)-U_\s(0,0)\geq \frac{c}{\lambda}.$$
(which must exist, since $\lim_{\s\to 0}U_\s(1,1)-U_\s(0,0)> c/\lambda $ by hypothesis.) By continuity, we thus have $h(\tilde \s)<0$. On the other hand, taking the limit as $\s\to \infty$, the difference $U_\s(1,1)-U_\s(0,0)$ remains bounded, and thus $\lim_{\s\to \infty} h(\s)>0$. By continuity, there thus must exist a value $\s^*$ such that $h(\s^*)=0$, which establishes the claim. 
\end{proof}

\end{appendix}

\end{document}